\documentclass[journal]{IEEEtran}
\usepackage{cite}
\usepackage{color,soul}
\usepackage{gensymb}
\usepackage{dsfont}

\ifCLASSINFOpdf
  \usepackage[pdftex]{graphicx}
  \graphicspath{{../pdf/}{../jpeg/}}
  \DeclareGraphicsExtensions{.pdf,.jpeg,.png}
\else
  \usepackage[dvips]{graphicx}
  \graphicspath{{../eps/}}
  \DeclareGraphicsExtensions{.eps}
\fi
\usepackage[cmex10]{amsmath}

\usepackage{amssymb}
{}

\usepackage{euscript}

\usepackage{multirow}
\usepackage{algorithm}
\usepackage{algpseudocode}
\usepackage{diagbox}
\usepackage{physics}
\usepackage{tikz}
\usepackage{comment}

\ifCLASSOPTIONcompsoc
  \usepackage[caption=false,font=normalsize,labelfont=sf,textfont=sf]{subfig}
\else
  \usepackage[caption=false,font=footnotesize]{subfig}
\fi
\usepackage{url}

\begin{document}
%
\title{A Data-Driven Framework for Assessing Cold Load Pick-up Demand in Service Restoration}
%
%
%

\author{Fankun Bu,~\IEEEmembership{Student Member,~IEEE,}
        Kaveh Dehghanpour,~\IEEEmembership{Member,~IEEE,}
        Zhaoyu~Wang,~\IEEEmembership{Member,~IEEE,}
        and~Yuxuan Yuan,~\IEEEmembership{Student Member,~IEEE}

\thanks{This work is supported by the Advanced Grid Modeling Program at the U.S. Department of Energy Office of Electricity under DE-OE0000875.

F. Bu, K. Dehghanpour, Z. Wang, and Y. Yuan are with the Department of
Electrical and Computer Engineering, Iowa State University, Ames, IA 50011
USA (e-mail: fbu@iastate.edu; wzy@iastate.edu).

}}

%
%

\markboth{}%
{Shell \MakeLowercase{\textit{et al.}}: Bare Demo of IEEEtran.cls for Journals}
%



\maketitle

\begin{abstract}
Cold load pick-up (CLPU) has been a critical concern to utilities. Researchers and industry practitioners have underlined the impact of CLPU on distribution system design and service restoration. The recent large-scale deployment of smart meters has provided the industry with a huge amount of data that is highly granular, both temporally and spatially. In this paper, a data-driven framework is proposed for assessing CLPU demand of residential customers using smart meter data. The proposed framework consists of two interconnected layers: 1) At the feeder level, a nonlinear auto-regression model is applied to estimate the diversified demand during the system restoration and calculate the CLPU demand ratio. 2) At the customer level, Gaussian Mixture Models (GMM) and probabilistic reasoning are used to quantify the CLPU demand increase. The proposed methodology has been verified using real smart meter data and outage cases.  
\end{abstract}


%
\IEEEpeerreviewmaketitle


\section{Introduction}
%
%
%
%
\IEEEPARstart{C}{old} load pick-up (CLPU) is a challenging issue in electric power industry \cite{model_physical_2, model_mathe_1, evaluation_2}. The CLPU demand is an increased load at service restoration phase due to the loss of load diversity. During normal system operation, the on-off switching cycles of thermostatically controlled loads (TCL) within a population of customers take place independently because of the heterogeneity of appliances and the diversity of customer behaviors. However, immediately after a long power outage in the restoration phase, the switching cycles of TCLs will coincide and become highly correlated for a period of time. This phenomenon is the main reason for the abnormal level of demand due to the temporary lack of diversity. In this paper, the term ``CLPU demand" refers to this undiversified load during the restoration. 

For feeders with high penetration of TCLs, CLPU can have serious consequences, such as restoration failure \cite{restore_1,restore_2,restore_3,restore_4,restore_5}, transformers aging \cite{transformer_1, transformer_2}, transformer overloading \cite{design_1}, and unacceptable voltage drops \cite{design_3}. CLPU demand can continue several minutes to even several hours after extreme weather conditions \cite{model_mathe_1}. Hence, it is necessary to quantify the impact of CLPU on distribution system design and restoration. To achieve this goal, the primary task is to quantify the deviation of CLPU demand from normal (diversified) load in historical outage cases. This will help the utilities to extract useful information for future service restorations.  

Previous papers have mainly focused on model-driven methods for CLPU demand assessment. In \cite{model_physical_2}, a physical model was built for simulating steady and transient response of thermostatically-controlled residential electric space heating devices. Based on the developed model, the aggregate impacts of space heaters on feeder-level CLPU demand were analyzed. In \cite{model_physical_1}, a simple and practical model was developed to represent temperature dynamics in a house with a thermostatically-controlled heater/air-conditioner. The model can be used in load management and aggregate CLPU impact evaluation. In \cite{model_physical_4}, similar groups of elementary component load models were built and the load models in the same group were aggregated by using statistical techniques to simulate CLPU. In \cite{model_physical_5}, a multi-state physical load model was developed to capture the behavior of end-use loads. Besides using air temperature as a control signal, other variables, such as price, can also be integrated into the model. In \cite{evaluation_2}, the developed model in \cite{model_physical_5} was used to account for the multi-state operation of residential heat pumps. This model was then employed to estimate the magnitude and duration of CLPU demand. In \cite{evaludation_1}, CLPU demands of seven houses with different types of electric heating equipment were measured, and field studies were also performed for load restoration process. Although \cite{evaludation_1} used field measurements to analyze CLPU demand, the employed dataset was procured from the measurements of only a limited number of residential customers, and it fails to employ estimation methods to capture the variations in the expected diversified demand at the time of restoration. Thus, previous works are largely dependent on detailed dynamic modeling of residential/commercial appliances.

While model-driven methods for CLPU demand evaluation offer benefits, such as physical interpretability and cost-efficiency, their disadvantages cannot be ignored. Residential loads depend on many factors, such as the types of appliances, the states of appliances, customer behaviors, and house thermal resistance and capacitance. Therefore, to accurately model a house load, enough detailed information should be collected, which is very challenging to accomplish for utilities in practice. This lack of such detailed information can lead to considerable modeling bias. On the other hand, in the past decade, smart meter data with high temporal-spatial granularity has become widely available to utilities \cite{the_survey}, which provides an opportunity to address the shortcomings of previous model-based approaches. Specifically, the impact of all the aforementioned unknown factors can be reflected in the smart meter data. For example, a larger thermal inertia value results in a decrease in the rate of change of indoor temperature that leads to less power consumption at the time of restoration, which is then measured by the smart meters. Hence, utilities can obtain CLPU information from smart meter data with high fidelity, instead of relying on model-driven techniques, which need detailed explicit knowledge of thermal parameters and appliances' information. Considering this, in this paper, we will develop a data-driven framework to assess feeder-level CLPU demand ratio and customer-level CLPU demand increase. Compared with the model-driven methods, the framework can provide an approach for assessing CLPU demand without the need for developing specific load models.

The main contribution of this work is to propose a framework for assessing CLPU demand at the time of restoration, which is used for developing statistical CLPU models. The proposed framework consists of two layers: 1) At the feeder level, the ratio of the undiversified CLPU demand to the estimated diversified demand is obtained. To achieve this, a least squares support vector machine (LS-SVM) auto-regression model is employed to estimate the diversified demand under the assumption that the outage has not happened. Then the CLPU demand ratio is calculated by dividing the actual undiversified CLPU demand at the time of restoration by the estimated diversified demand. Finally, a CLPU ratio regression model is developed based on the obtained historical CLPU ratios under different outage duration and ambient temperatures. The developed regression model can be used for predicting how the load behaves under new and previously unseen outage cases. Therefore, one innovative aspect of this paper is using a load estimation technique to assess feeder-level CLPU demand, which has not been applied in previous papers regarding CLPU demand evaluation. 2) At the customer level, a novel CLPU demand increase assessment approach is proposed. Gaussian mixture models (GMM) are applied to devise a probabilistic technique towards constructing marginal probability density functions (PDF) of customer demand increase due to CLPU. Then the customer CLPU demand increase is analyzed statistically for a set of customers to evaluate the loss of load diversity. The performance of the developed framework is verified using real smart meter data from three Midwest U.S. utilities. We have also shown that using the proposed approach the PDF of demand increase due to CLPU can be estimated for any group of customers, which can provide an invaluable guideline for designing sequential restoration plans for distribution systems.

The rest of the paper is organized as follows: Section \ref{sec:overall} introduces the proposed framework of CLPU demand assessment and the real dataset. Section \ref{sec:feeder} presents the procedure of feeder CLPU demand ratio evaluation. In Section \ref{sec:customer}, the procedure of assessing customer CLPU demand increase is presented. In Section \ref{sec:casestudy}, outage case studies are analyzed. Section \ref{sec:conclusion} concludes the paper.

\section{Proposed CLPU Demand Assessment Framework and Real Dataset Description}\label{sec:overall}
\begin{figure}
\centering
\includegraphics[width=0.38\textwidth]{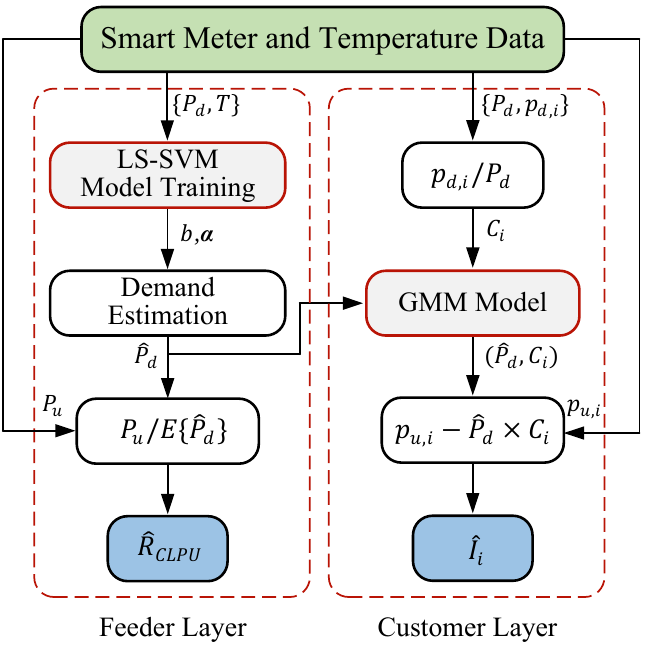}
\caption{CLPU demand assessment framework.}
\label{fig:Flow_Chart}
\end{figure}

This section presents a high-level overview of the proposed framework for feeder-level and customer-level CLPU demand assessment. We will also describe the real smart meter dataset available to us. Assuming customers' smart meter data during normal and restoration conditions is available to utilities, then the feeder-level demand can be obtained by aggregating time-aligned customer-level demand. The overall framework is shown in Fig. \ref{fig:Flow_Chart}. 

\subsection{Feeder Layer}
\begin{figure}
\centering
\includegraphics[width=0.8\linewidth]{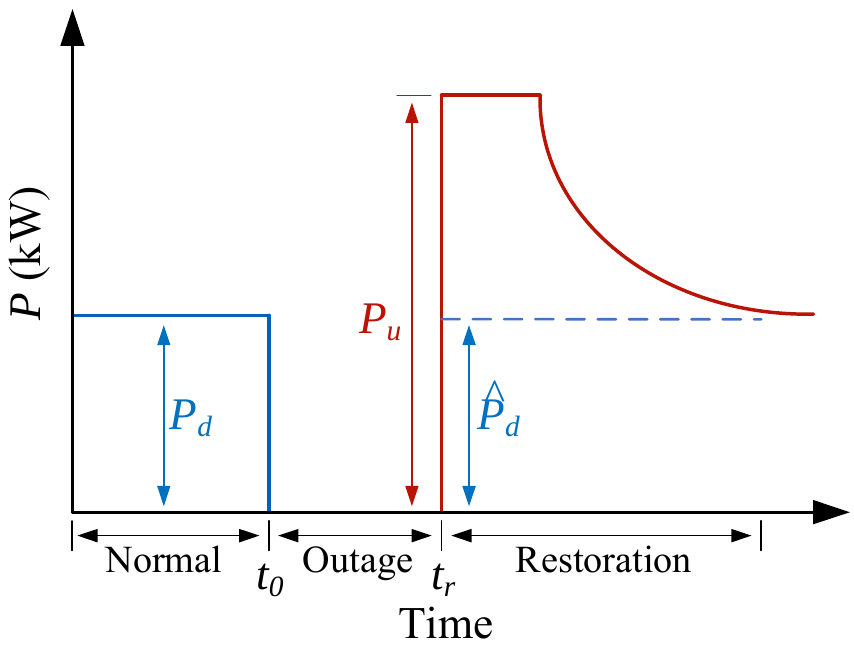}
\caption{Typical CLPU demand curve.}
\label{fig:expon_model}
\end{figure}

The objective of feeder-level CLPU demand assessment is to obtain the CLPU demand ratio, $R_{CLPU}$, defined as follows:
\begin{equation}  \label{eq:CLPU_ratio}
R_{CLPU}=\frac{P_u}{P_d}
\end{equation}
where, $P_u$ is the undiversified feeder demand at the time of restoration, $t_r$, after outage occurrence at $t_0$, and $P_d$ is the diversified feeder demand at time $t_r$. However, as shown in Fig. \ref{fig:expon_model}, the actual measured feeder demand at the time of restoration is undiversified due to CLPU. This implies that $P_d$ at time $t_r$ cannot be directly obtained from smart meter data, and needs to be estimated based on the demand data history during normal system operation.  Hence, the first task at this level is to estimate $P_d$ at the time of restoration. For this, we design a machine learning technique which is based on a nonlinear auto-regression model with exogenous input (NARX) and is implemented using LS-SVM. The LS-SVM is trained using historic $P_d$ and temperature data to continuously predict the future diversified demand. Since the actual feeder load at the time of restoration is undiversified due to CLPU, the outcome of the machine learning model is the estimated diversified feeder demand \textit{if the outage did not happen}. Therefore, taking the estimation residuals into account, the estimated diversified feeder demand $\hat{P}_d$ at time $t_r$ is modeled as follows:
\begin{equation}  \label{eq:hat_P}
\hat{P}_d=P_d+\varepsilon_{t_r}
\end{equation}
where, $\varepsilon_{t_r}$ denotes the machine learning estimation residual. Assuming that the residual follows a Gaussian distribution, namely $\varepsilon_{t_r}\sim \mathcal{N}(0,\sigma^{2})$, with $\sigma$ defining the machine learning framework uncertainty, the estimated diversified demand also follows a Gaussian distribution \cite{LSSVM_2}:
\begin{equation}  \label{eq:hat_P_estimation}
\hat{P}_{d}\sim \mathcal{N}(P_d,\sigma^{2})
\end{equation}

Therefore, the CLPU demand ratio can be estimated as follows:
\begin{equation}  \label{eq:CLPU_ratio_hat}
\hat{R}_{CLPU}=\frac{P_u}{E\{\hat{P}_d\}}
\end{equation}
where, $E\{\cdot\}$ is the empirical averaging operator.

\subsection{Customer Layer}
The objective of this level is to construct the marginal PDF of individual customer demand increase due to CLPU at the time of restoration, denoted as $I_i$ for the $i^{th}$ customer and defined as follows:
\begin{equation}  \label{eq:demand_incres_def}
I_{i}=p_{u,i}(t_r)-p_{d,i}(t_r)
\end{equation}
where, $p_{u,i}(t_r)$ is the actual customer demand corresponding to the undiversified feeder load, and $p_{d,i}(t_r)$ is the customer demand at the time of restoration \textit{if the outage did not happen}. Similar to the variable $P_d$ in equation \eqref{eq:CLPU_ratio}, since $p_{d,i}(t_r)$ is unknown and cannot be measured by smart meters, it needs to be estimated. However, compared to feeder demand, customer demand can be much more volatile. Considering this volatility, instead of directly estimating $p_{d,i}(t_r)$ for each customer, this paper adopts a probabilistic learning approach to construct the marginal PDF of the estimated customer demand at time $t_r$ using the obtained $\hat{P}_d$ for the feeder. Based on the demand data history from smart meters, a \textit{contribution factor} is defined for the $i^{th}$ customer, denoted as $C_i$, which determines the customer contribution to feeder demand ($P_d$). Note that $C_i$ is obtained in normal state (without outage) when $p_{d,i}(t)$ can be measured directly by smart meters, as follows: 
\begin{equation}  \label{eq:CF_definition}
C_{i}(t)=\frac{p_{d,i}(t)}{\sum_{j=1}^{M}p_{d,j}(t)}=\frac{p_{d,i}(t)}{P_d(t)} \quad i=1,\cdots,M.
\end{equation}
where, $M$ is the total number of customers connected to the feeder. Hence, during the normal operation, an individual customer demand can be determined as $p_{d,i}(t)=P_d(t)C_i(t)$. Noting the dependency of $p_{d,i}(t)$ on both $P_d(t)$ and $C_i(t)$, to obtain the marginal PDF of $\hat{p}_{d,i}(t_r)$, the joint PDF of the estimated diversified feeder demand ($\hat{P}_d(t)$) and the contribution factor at the time of restoration is constructed. This joint PDF is determined using a GMM technique, which employs past customer demand measurements and the corresponding estimated diversified feeder demand. It will be shown that a nonlinear transformation of this joint PDF can be used to obtain the marginal PDF for $\hat{p}_{d,i}(t_r)$. The CLPU demand increase for the $i^{th}$ customer at the time of restoration, $\hat{I}_i$, is estimated as:
\begin{equation}  \label{eq:demand_incres_def}
\hat{I}_{i}=p_{u,i}(t_r)-\hat{p}_{d,i}(t_r)
\end{equation}
Note that given the obtained marginal PDF for $\hat{p}_{d,i}$, equation \eqref{eq:demand_incres_def} also leads to a marginal PDF for the individual customer demand increase.


\subsection{Real Dataset Description}

The available smart meter data history contains three U.S. Midwest utilities' energy consumption data (kWh) of over 10,000 residential customers with a 15-minute time resolution, and the time range is about four years. This data includes time stamps which have been used for customer-level demands' time-alignment. When an outage occurs, the meter will keep a record of outage start time, end time, and the associated energy consumption readings. The ambient temperature data is obtained from the National Oceanic and Atmospheric Administration (NOAA) database \cite{NOAAwebsite} , and is time-aligned with smart meter data.

\section{Feeder-Level Diversified Demand Estimation}\label{sec:feeder}  
In this section, a LS-SVM regression model is developed to estimate the diversified feeder demand, $\hat{P}_d$, at the time of restoration, which is then used to determine the CLPU demand ratio, $R_{CLPU}$. 

The LS-SVM is based on a support vector margin maximization process to minimize the machine learning structural risk function. This regression model has many advantages, including good generalization capability and low susceptibility to local minima \cite{LSSVM_1, LSSVM_2}, and has been employed in distribution systems \cite{LSSVM_3, LSSVM_4}. In demand estimation, the selection of explanatory variables is critical. The feeder demand at a certain time is primarily affected by the temperature at that time and is highly correlated with previous demand samples within a certain time period \cite{LSSVM_2}. Demand also changes with seasons and days of week (working day vs non-working day). To capture seasonal and daily demand diversity, the dataset is divided across seasons and working/non-working days, respectively. The explanatory variable $\mathbf{x}(t)\in\mathbb{R}^n$, which acts as the input to the demand estimation model, is built as follows:
\begin{equation}  \label{eq:LSSVM_eq1}
\mathbf{x}(t)=[P_d(t-1),\cdots,P_d(t-n_{lag}),T(t)]^\mathsf{T}
\end{equation}
where, $P_d(t-i)$ is the feeder demand at time $t-i$, $n_{lag}$ is the maximum time lag, and $T(t)$ is the ambient temperature at time $t$. Therefore, using this explanatory variable, feeder demand at time $t$ can be expressed as:
\begin{equation}  \label{eq:LSSVM_eq2}
P_d(t)=\boldsymbol\omega^\mathsf{T}\boldsymbol\varphi(\mathbf{x}(t))+b+\varepsilon_{t}
\end{equation}
where, $\boldsymbol{\omega}\in\mathbb{R}^{n_h}$ and $b\in\mathbb{R}$ represent regression model parameters, $\boldsymbol{\varphi}(\cdot):\mathbb{R}^n\to\mathbb{R}^{n_h}$ is a mapping function, transforming low dimensional input vector $\mathbf{x}(t)$ into a high dimensional vector $\boldsymbol\varphi(\mathbf{x}(t))$, and $\varepsilon_{t}$ is a normally distributed random variable representing the estimation residual.

Given the current feeder demand and temperature values, together with the past feeder demand samples with certain time lags, a training set of size $N_{tr}$ can be developed, $\mathbf{D}_{tr} = \{\mathbf{x}(t) ,  \mathbf{P}_d(t)\}_{t=1}^{N_{tr}}$. To obtain the optimal values of learning parameters, a structural risk function, $J$, is formulated and minimized with respect to $\boldsymbol\omega$, $b$, and $\varepsilon_{t}$, over the training set. This optimization is formulated as follows:
\begin{equation}  \label{eq:LSSVM_eq3}
\begin{split}
&\min_{\boldsymbol{\omega},b,\varepsilon_{t}}J=\frac{1}{2}\boldsymbol{\omega}^\mathsf{T}\boldsymbol{\omega}+\gamma\sum_{t=1}^{N_{tr}}\varepsilon_{t}^2\\
s.t.\ \ P_d(t)&=\boldsymbol\omega^\mathsf{T}\boldsymbol\varphi(\mathbf{x}(t))+b+\varepsilon_{t},\quad t=1,\cdots,N_{tr}.
\end{split}
\end{equation}
where, $\gamma$ is a regularization constant to prevent overfitting. To solve this optimization problem, the Lagrangian, $\mathcal{L}$, is constructed as a function of regression parameters:
\begin{multline}  \label{eq:LSSVM_eq5}
\mathcal{L}(\boldsymbol\omega,b,\varepsilon_{t};\boldsymbol{\alpha})=J(\boldsymbol{\omega},b,\varepsilon_{t})-\\
\sum_{t=1}^{N_{tr}}\alpha_{t}(\boldsymbol{\omega}^{T}\boldsymbol\varphi(\mathbf{x}_{t})+b+\varepsilon_{t}-P_d(t))
\end{multline}
where, $\alpha_{t}$'s are Lagrange multipliers. The optimality conditions are obtained by solving $\nabla_{(\boldsymbol{\omega},b,\varepsilon_{t})}\mathcal{L}=0$, as follows:
\begin{equation}  \label{eq:LSSVM_eq6}
\begin{cases}
\frac{\partial \mathcal{L}}{\partial \boldsymbol\omega}=0 \to \boldsymbol\omega = \displaystyle \sum_{t'=1}^{N_{tr}}\alpha_{t'}\boldsymbol\varphi(\mathbf{x}(t')) \\
\frac{\partial \mathcal{L}}{\partial b}=0  \to  \displaystyle \sum_{t=1}^{N_{tr}}\alpha_{t}=0\\
\frac{\partial \mathcal{L}}{\partial \varepsilon_{t}}=0  \to  \alpha_{t}=\gamma\varepsilon_{t}\\
\frac{\partial \mathcal{L}}{\partial \alpha_{t}}=0  \to  P_d(t)=\boldsymbol\omega^\mathsf{T}\boldsymbol\varphi(\mathbf{x}(t))+b+\varepsilon_{t}
\end{cases}
\end{equation}
Combining equations (\ref{eq:LSSVM_eq2}) and \eqref{eq:LSSVM_eq6}, $\boldsymbol{\omega}$ can be eliminated from the regression model as shown below:
\begin{equation}  \label{eq:LSSVM_eq7}
P_d(t)=\sum_{t'=1}^{N_{tr}}\alpha_{t}\boldsymbol\varphi(\mathbf{x}(t'))^\mathsf{T}\boldsymbol{\varphi}(\mathbf{x}(t))+b+\frac{\alpha_{t}}{\gamma} \quad t=1,\cdots,N_{tr}.
\end{equation}
The term $\boldsymbol\varphi(\mathbf{x}(t'))^\mathsf{T}\boldsymbol{\varphi}(\mathbf{x}(t))$ in equation (\ref{eq:LSSVM_eq7}) can be represented by a \textit{kernel function}, $K(.,.)$, as follows:
\begin{equation}    \label{eq:LSSVM_eq8} 
K(\mathbf{x}(t'),\mathbf{x}(t)) = \boldsymbol\varphi(\mathbf{x}(t'))^\mathsf{T}\boldsymbol{\varphi}(\mathbf{x}(t)) \quad t',t=1,\cdots,N_{tr}.
\end{equation}
In this paper, a Gaussian kernel is employed to replace the dot product in equation (\ref{eq:LSSVM_eq8}):
\begin{equation}    \label{eq:LSSVM_eq9} 
K(\mathbf{x}(t'),\mathbf{x}(t)) = \exp(-\frac{||{\mathbf{x}(t')-\mathbf{x}(t)}||^2}{\sigma^2}) \quad t',t=1,\cdots,N_{tr}.
\end{equation}

Note that equations (\ref{eq:LSSVM_eq6}) and (\ref{eq:LSSVM_eq7}) yield a set of linear equations, from which the machine learning parameters, $b$ and $\pmb{\alpha}$ are obtained for the given training set $\mathbf{D}_{tr}$:
\begin{equation}   \label{eq:LSSVM_eq11} 
\left[
\begin{array}{ccc}
b  \\
\boldsymbol{\alpha}
\end{array}
\right] 
=
\left[
\begin{array}{ccc}
0 & \mathbf{1}^\mathsf{T} \\
\mathbf{1} & \Omega+\frac{1}{\gamma}\mathbf{I} 
\end{array}
\right]^{-1} 
\left[
\begin{array}{ccc}
0 \\
\mathbf{P}_d
\end{array}
\right] 
\end{equation}
where, $\mathbf{P}_d=[P_d(1),\cdots,P_d(N_{tr})]^\mathsf{T}$, $\mathbf{1}=[1,\cdots,1]^\mathsf{T}$, $\mathbf{I}$ is the identity matrix, $\boldsymbol{\alpha}=[\alpha_{1},\cdots,\alpha_{N_{tr}}]^\mathsf{T}$, and the entries of the kernel matrix, $\Omega$, are determined as follows:
\begin{equation}    \label{eq:LSSVM_eq10} 
\Omega_{t't}=K(\mathbf{x}(t'),\mathbf{x}(t)) \quad t',t=1,\cdots,N_{tr}.
\end{equation}
It should be noted that $b$ and $\boldsymbol\alpha$ can have different values as the values of input parameters, $n_{lag}$, $\sigma$, and $\gamma$, change. To tune the regression model with respect to input parameters, k-fold cross-validation is performed. Moreover, mean absolute percentage error (MAPE) is adopted as the criteria for evaluating the performance of the regression model. After completing the cross-validation and training procedures, which optimize the input variables and learning parameters, the estimation accuracy of the regression model is evaluated on a test set, $\mathbf{D}_t$, of size $N_t$.

The critical step in calculating feeder CLPU demand ratio $R_{CLPU}$ is to estimate the diversified feeder demand at the time of restoration, $t_r$. Similar to equation (\ref{eq:LSSVM_eq1}), the explanatory variable at time $t_r$ is obtained as follows:
\begin{equation}  \label{eq:LSSVM_xt0}
\mathbf{x}(t_r)=[P_d(t_r-1),\cdots,P_d(t_r-n_{lag}^*),T(t_r)]^\mathsf{T}
\end{equation}
where, $n_{lag}^*$ is the optimal time lag. Using this explanatory variable, the estimated diversified feeder demand at time $t_r$ is determined based on the trained LS-SVM model, as follows:
\begin{equation}      \label{eq:LSSVM_Pt0} 
\hat{P}_d(t_r)=\sum_{t=1}^{N_{tr}}\alpha_{t}^*K(\mathbf{x}(t),\mathbf{x}(t_r))+b^*
\end{equation}
where, $\alpha_{t}^*$ and $b^*$ are the optimal machine learning parameters.

Hence, $R_{CLPU}$ is obtained by dividing the undiversified feeder restoration demand by the empirically averaged estimated diversified feeder demand at time $t_r$, as shown in equation (\ref{eq:CLPU_ratio_hat}). Note that the empirical averaging process is performed considering the estimation residual distribution obtained from a test set during normal system operation. An algorithmic overview of LS-SVM model for assessing $R_{CLPU}$ is summarized in Algorithm \ref{alg:LSSVM}.

\begin{algorithm}
\caption{LS-SVM}\label{alg:LSSVM}
\begin{algorithmic}[1]
\State {Split demand and temperature data into two parts: training/validation set, and test set}
\Procedure{Training/Validation}{}
    \State {Select initial $n_{lag},\sigma,\gamma$}
    \State {$\mathbf{D}_{tr} \gets \left\{\mathbf{x}(t),P_d(t)\right\}_{t=1}^{N_{tr}}$}
    \State {$\Omega \gets K(\mathbf{x}(t'),\mathbf{x}(t)) \quad t',t=1,\cdots,N_{tr}.$}
    \State {$\mathbf{P}_d \gets [P_d(1),\cdots,P_d(N_{tr})]^\mathsf{T}$}
    \State {Solve equation (\ref{eq:LSSVM_eq11}) to obtain $\alpha_t$'s and $b$}
    \State {$\hat{P}_d(t') \gets \sum_{t=1}^{N_{tr}}\alpha_{t}K(\mathbf{x}(t),\mathbf{x}(t'))+b$}
    \State {Compute validation MAPE}
    \State {Change $n_{lag},\sigma,\gamma$, do Step 3 to 9, optimize parameters, $n_{lag}^*,\sigma^*,\gamma^*, b^*, \boldsymbol{\alpha}^*$}
\EndProcedure

\Procedure{Testing}{$n_{lag}^*,\sigma^*,\gamma^*, b^*, \boldsymbol{\alpha}^*$}
    \State {$\mathbf{D}_{t} \gets \left\{\mathbf{x}(t),P_d(t)\right\}_{t=1}^{N_{t}}$}
     \State {$\hat{P}_d(t)=\sum_{t'=1}^{N_{tr}}\alpha_{t'}^*K(\mathbf{x}(t'),\mathbf{x}(t))+b^*$}
    \State {Compute test MAPE}
\EndProcedure

\Procedure{Demand Estimation}{$n_{lag}^*,\sigma^*,\gamma^*, b^*, \boldsymbol{\alpha}^*$}
    \State {$\mathbf{x}(t_r) \gets [P_d(t_r-1),\cdots,P_d(t_r-n_{{lag}}^*),T(t_r)]^\mathsf{T}$}
    \State {$\hat{P}_d(t_r)\gets \sum_{t=1}^{N_{tr}}\alpha_{t}^*K(\mathbf{x}(t),\mathbf{x}(t_r))+b^*$}
\EndProcedure
\State {$\hat{R}_{CLPU} \gets P_u/E\{\hat{P}_d(t_r)\}$}
\end{algorithmic}
\end{algorithm}

\section{Customer Demand Increase Estimation}\label{sec:customer}

\begin{figure}[h]
\centering
\includegraphics[width=0.8\linewidth]{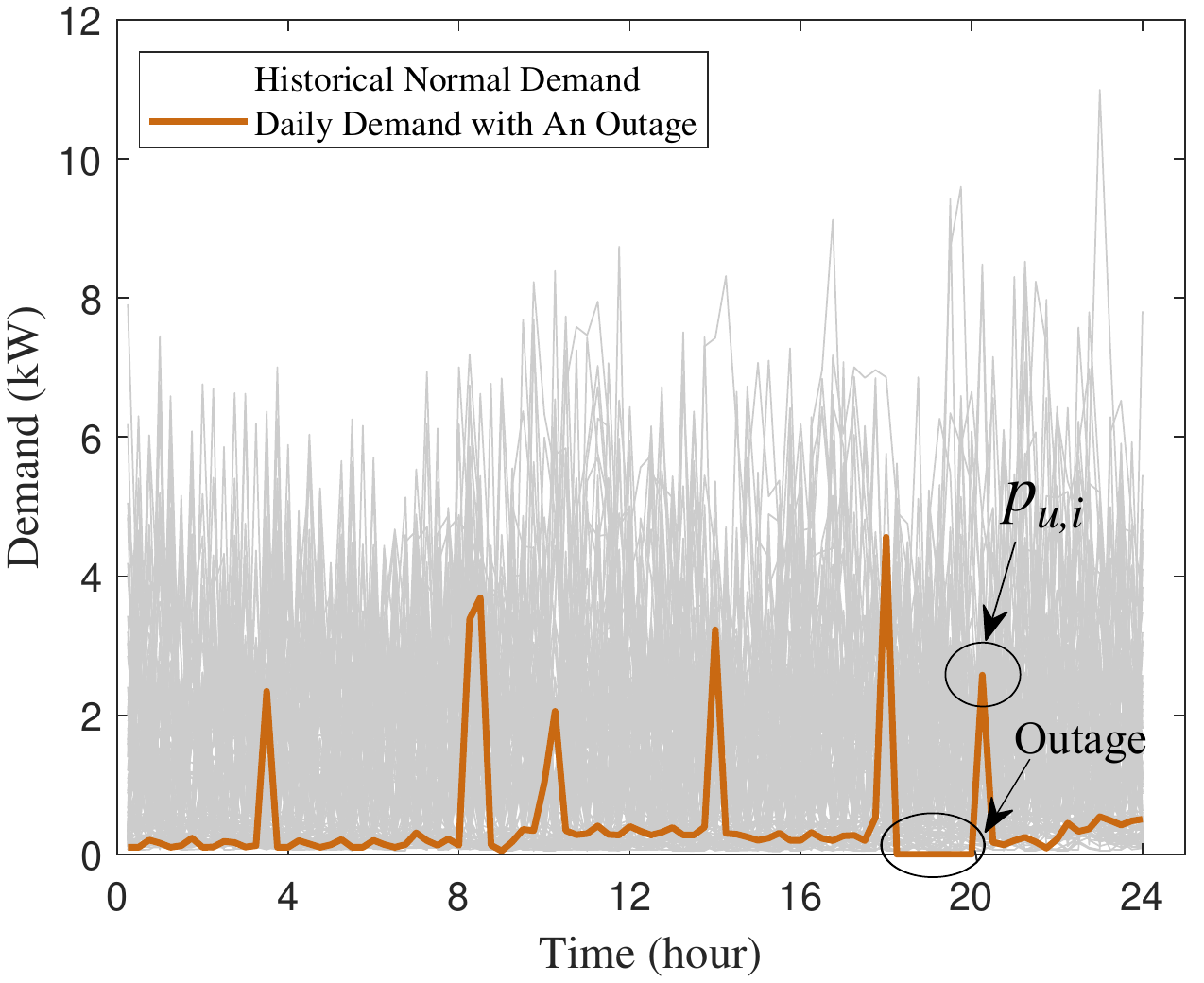}
\caption{Customer daily demand curves.}
\label{fig:customer_demand_curve}
\end{figure}

\begin{figure}
      \centering
      \includegraphics[width=1\columnwidth]{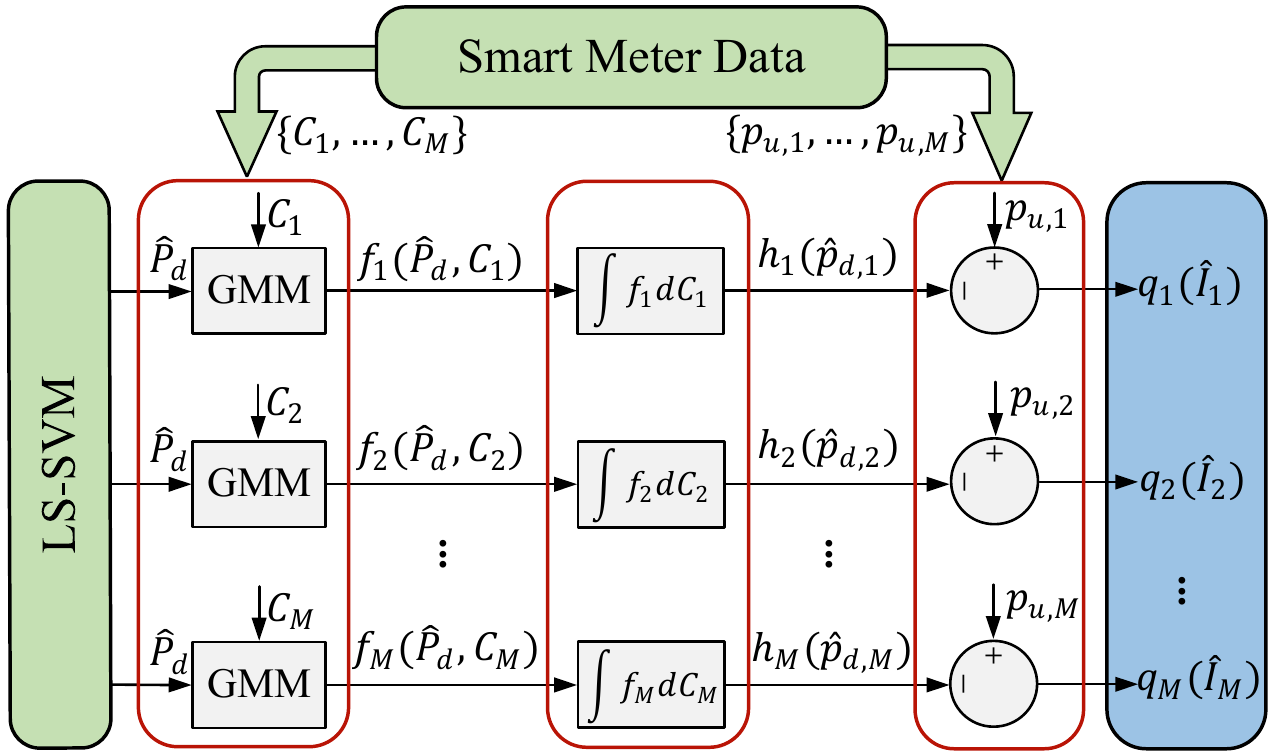}
\caption{Overall structure of the customer CLPU demand increase assessment.}
\label{fig:GMM_overall}
\end{figure}

Although the aggregate residential demand at feeder level can be estimated with satisfactory accuracy, individual customer consumption can be quite stochastic \cite{VSTLF_2}. Fig. \ref{fig:customer_demand_curve} shows the daily demand curves of a single residential customer with and without outage, with 15-minute resolution during a season. The gray curves are the historical daily demand data during normal operation and the red curve denotes the demand data in one day with an outage. The spikes represent the semi-periodic on/off cycling of customer appliances, which are captured by the smart meter. It can also be seen that customer demand at the interval [17.75 h, 20 h] equals zero, indicating an outage during this period. The plot shows that daily load curves do not present obvious cyclic behavior in contrast with the feeder demand. Also, customer peak demand at the time of restoration, $p_{u,i}$, is not necessarily larger than normal demand. Hence, considering the volatility of customer demand, in this section a probabilistic method is proposed to determine $\hat{I}_i$. The overall structure of customer-level CLPU demand assessment is shown in Fig. \ref{fig:GMM_overall}.

As discussed in Section II, the estimated demand for the $i^{th}$ customer at the time of restoration is obtained as follows:
\begin{equation}  \label{eq:cust_3}
\hat{p}_{d,i}(t_r)=\hat{P}_d(t_r)C_i  \quad i=1,\cdots,M.
\end{equation}

The difficulty in equation (\ref{eq:cust_3}) is to compute the product of the two random variables $\hat{P}_{d}$ and $C_{i}$. It is shown in \cite{product_two_variable}, that using the joint PDF of two dependent random variables, the marginal PDF of their product can be obtained using a nonlinear transformation. Hence, denoting the joint PDF of $\hat{P}_{d}$ and $C_{i}$ by $f_i(\hat{P}_d,C_i)$, the marginal PDF of estimated customer demand $\hat{p}_{d,i}$ at time $t_r$ is obtained as follows \cite{product_two_variable}:
\begin{equation}  \label{eq:cust_5}
h_i(\hat{p}_{d,i})=\int_{0_+}^{1}f_i\big(\frac{\hat{p}_{d,i}}{C_i}, C_i\big) \frac{1}{C_i} dC_i
\end{equation}

Therefore, the first step in calculating equation \eqref{eq:cust_5} is to obtain $f_i(\hat{P}_d,C_i)$ for each customer. To do this, a probabilistic technique is employed using GMMs. GMM is a parametric model, which approximates arbitrary PDFs as weighted summation of Gaussian density components. GMM has been previously applied in distribution systems studies for modeling the stochasticity of load and the uncertainty of distribution system state estimators\cite{GMM_1, GMM_2}. In this paper, we propose using GMM to model the joint PDF of customer contribution and the estimated diversified feeder demand. Thus, based on the estimated diversified feeder demand, $\hat{P}_{d}$, and contribution factor, $C_{i}$, for the $i^{th}$ customer, the GMM approximation model, which is composed of $S_{i}$ Gaussian components, can be expressed as follows:
\begin{equation}  \label{eq:GMM_1}
l(\mathbf{z}|\boldsymbol{\lambda})=\displaystyle\sum_{j=1}^{S_{i}}w_{j}g(\mathbf{z}|\boldsymbol{\mu}_{j},\boldsymbol{\Sigma}_{j})
\end{equation}
where, $\mathbf{z}$ is a two-dimensional continuous-valued vector defined as $\mathbf{z}=[\hat{P}_d,C_i]$, $w_j$'s are the mixture weights corresponding to multi-variate Gaussian components $g(\mathbf{z}|\boldsymbol{\mu}_{j},\boldsymbol{\Sigma}_{j})$, which satisfy $\sum_{j=1}^{S_{i}}w_{j}=1$. Thus, each component is a bi-variate Gaussian function defined as:
\begin{multline}  \label{eq:GMM_2}
g(\mathbf{z}|\boldsymbol{\mu}_{j},\boldsymbol{\Sigma}_{j})=\frac{1}{(2\pi)|\Sigma_{j}|^{1/2}}\\
\exp\Big\{-\frac{1}{2}(\mathbf{z}-\boldsymbol{\mu}_{j})^\top\boldsymbol{\Sigma}_{j}^{-1}(\mathbf{z}-\boldsymbol{\mu}_{j})\Big\}
\end{multline}
where, $\boldsymbol{\mu}_{j}$ and $\boldsymbol{\Sigma}_{j}$ are the component mean vector and covariance matrix, respectively. Also,  $\boldsymbol{\lambda}=\{w_{j},\boldsymbol{\mu}_{j},\boldsymbol{\Sigma}_{j}\}$, is the collection of parameters of the GMM model that have to be learned.

With respect to GMM approximation, the model training objective is to obtain the optimal parameter collection $\boldsymbol{\lambda}^*$ that best matches the distribution of the given training set. The training set is composed of $C_i$ data history and $\hat{P}_d$ samples. The most well-established method for GMM training is the maximum likelihood (ML) estimation \cite{GMM}. Given the training vectors $Z=\{\mathbf{z}_1,\cdots,\mathbf{z}_{N}\}$ with $N$ samples, the GMM likelihood can be written as:
\begin{equation}  \label{eq:GMM_3}
l(Z|\boldsymbol{\lambda})=\displaystyle \prod_{t=1}^{N}p(\mathbf{z}_{t}|\boldsymbol{\lambda})
\end{equation}
Generally, this non-linear function can be minimized iteratively with respect to $\pmb{\lambda}$ using expectation-maximization (EM) algorithm \cite{GMM}. Also, to tune the GMM models with respect to the component number, $S_i$, k-fold cross-validation is performed. To evaluate model performance with different learned $\boldsymbol{\lambda}$'s, the Bayesian information criterion (BIC) is employed.

Based on the optimal $\boldsymbol{\lambda}^*$ obtained from equation (\ref{eq:GMM_3}) and using equations (\ref{eq:GMM_1}) and (\ref{eq:GMM_2}), the joint PDF of estimated feeder demand $\hat{P}_d$ at time $t_r$ and contribution factor $C_i$ for the $i^{th}$ customer can be specifically written as,
\begin{equation}    \label{eq:cust_1}
f_i(\hat{P}_d,C_i)=\displaystyle \sum_{j=1}^{S_{i}}\omega_{j}g(\hat{P}_d,C_i)
\end{equation}
where,
\begin{multline}  \label{eq:cust_2}
g(\hat{P}_d,C_i)=\frac{1}{2\pi \sigma_{\hat{P}_d}^{(j)} \sigma_{C_i}^{(j)} \sqrt{1-\rho_{j}^{2}}} \exp \bigg(-\frac{1}{2(1-\rho_{j}^2)}\\
\Big[\frac{(\hat{P}_d-\mu_{\hat{P}_d}^{(j)})^2}{{\sigma_{\hat{P}_d}^{(j)}}^2} +\frac{(C_i-\mu_{C_i}^{(j)})^2}{{\sigma_{C_i}^{(j)}}^2} - \frac{2\rho_{j}(\hat{P}_d-\mu_{\hat{P}_d}^{(j)})(C_i-\mu_{C_i}^{(j)})}{\sigma_{\hat{P}_d}^{(j)}\sigma_{C_i}^{(j)}}   \Big] \bigg)
\end{multline}
where, $\mu_{\hat{P}_d}^{(j)}$, $\mu_{C_i}^{(j)}$, $\sigma_{\hat{P}_d}^{(j)}$, $\sigma_{C_i}^{(j)}$, and $\rho_{j}$ are the corresponding mean, variance, and correlation of $\hat{P}_d$ and $C_i$ for the $j$th component, respectively. Hence, substituting equation \eqref{eq:cust_1} into \eqref{eq:cust_5}, the marginal PDF of the estimated customer demand, $h_i(\hat{p}_{d,i})$, is obtained using numerical integration over the customer contribution factor variable. 

Finally, using equation \eqref{eq:demand_incres_def}, the marginal PDF of demand increase for the $i^{th}$ customer is constructed. Note that since $p_{u,i}$ is directly measured by the customer's smart meter at the time of restoration, it is treated as a constant value. Hence, the marginal PDF of $\hat{I}_i$, denoted as $q_i$ is obtained for each customer, as follows:
\begin{equation}    \label{eq:pdf_I}
q_i(\hat{I}_i)= h_i(p_{u,i} - \hat{I}_{i})
\end{equation}

\section{Case Study}\label{sec:casestudy}
Nineteen outage cases are observed for evaluating feeder-level CLPU demand ratio and post-outage customer-level CLPU demand increase. The case information is shown in Table \ref{tbl:outage_cases}. 
\begin{table}[h]
\begin{center}
\caption{Outage Case Information}\label{tbl:outage_cases}
\renewcommand{\arraystretch}{1.4}
\begin{tabular}{|c|c|c|}
	\hline
     Case & Outage Duration & Ambient Temperature\\ 
	\hline
    1 & 33 min & $38.5\ ^{\circ}{\rm C}$\\
	\hline
    2 & 91 min & $37.0\ ^{\circ}{\rm C}$\\  
	\hline
	3 & 128 min & $39.0\ ^{\circ}{\rm C}$\\  
	\hline
	4 & 136 min & $35.0\ ^{\circ}{\rm C}$\\
	\hline
	5 & 61 min & $31.5\ ^{\circ}{\rm C}$\\
	\hline
    6 & 83 min & $34.0\ ^{\circ}{\rm C}$\\
	\hline
    7 & 47 min & $35.5\ ^{\circ}{\rm C}$\\  
	\hline
	8 & 101 min & $30.0\ ^{\circ}{\rm C}$\\  
	\hline
	9 & 46 min & $25.5\ ^{\circ}{\rm C}$\\
	\hline
	10 & 77 min & $29.5\ ^{\circ}{\rm C}$\\
	\hline
    11 & 93 min & $38.5\ ^{\circ}{\rm C}$\\
	\hline
    12 & 41 min & $31.0\ ^{\circ}{\rm C}$\\  
	\hline
	13 & 65 min & $36.5\ ^{\circ}{\rm C}$\\  
	\hline
	14 & 48 min & $29.0\ ^{\circ}{\rm C}$\\
	\hline
	15 & 63 min & $38.0\ ^{\circ}{\rm C}$\\
	\hline
    16 & 118 min & $26.5\ ^{\circ}{\rm C}$\\  
	\hline
	17 & 107 min & $32.0\ ^{\circ}{\rm C}$\\  
	\hline
	18 & 85 min & $27.0\ ^{\circ}{\rm C}$\\
	\hline
	19 & 35 min & $29.5\ ^{\circ}{\rm C}$\\
	\hline
\end{tabular}
\end{center}
\end{table}

\subsection{Feeder CLPU Demand Ratio Estimation}
\subsubsection{CLPU Ratio Estimation and Regression Analysis}
\begin{figure}
\centering
\includegraphics[width=0.8\linewidth]{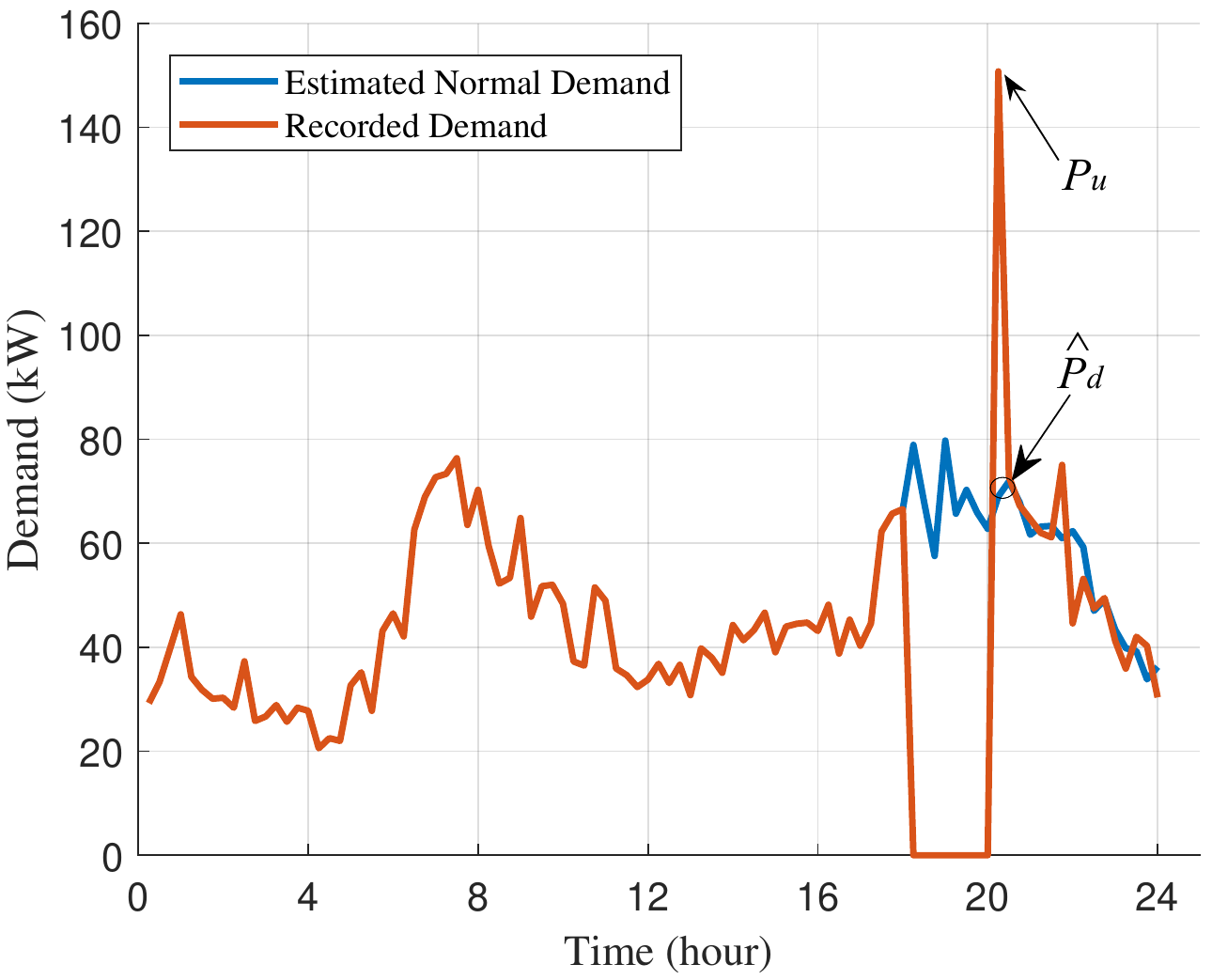}
\caption{Estimated diversified feeder demand curve and the recorded demand curve with an outage.}
\label{fig:feeder_curve}
\end{figure}

Feeder CLPU demand ratio is obtained by dividing the measured undiversified restoration demand by the estimated diversified demand at the time of restoration. As shown in Fig. \ref{fig:feeder_curve}, a demand overshoot occurs in the restoration phase, and the undiversified demand is significantly greater than the estimated diversified demand. Note that the undiversified CLPU demand (the spike labeled as $P_u$ in Fig. 5) is observed using smart meter data, and is not estimated. Also, it is observed that once the restoration phase is completed, the actual feeder demand drops back to the estimated diversified levels. This corroborates the accuracy of the LS-SVM framework.

\begin{table}[h]
\begin{center}
\caption{Feeder CLPU Demand Ratio}\label{tbl:CLPU_ratio}
\renewcommand{\arraystretch}{1.4}
\begin{tabular}{|c|c|c|c|c|c|c|}
	\hline
    \multicolumn{1}{|c|}{\multirow{2}{*}{Case}} & \multicolumn{2}{c|}{LS-SVM} & \multicolumn{2}{c|}{P-NARX} & \multicolumn{2}{c|}{ARX} \\
    \cline{2-7}
    \multicolumn{1}{|c|}{}&   Ratio & MAPE & Ratio & MAPE & Ratio & MAPE\\
	\hline
    1 & 1.41 & 5.45\% & 1.28 & 6.85\% & 1.37 & 6.38\% \\ 
	\hline
    2 & 1.63 & 10.89\% & 1.56 & 13.59\% & 1.55 & 13.70\% \\  
	\hline
	3 & 1.88 & 11.56\% & 1.83 & 13.70\%& 1.81 & 13.55\% \\ 
	\hline
	4 & 2.16 & 10.77\% & 2.25 & 12.30\%& 2.19 & 11.88\% \\ 
	\hline
    5 & 2.87 & 10.48\% & 2.97 & 13.53\% & 2.88 & 10.45\% \\ 
	\hline
    6 & 2.69 & 7.09\% & 2.72 & 5.62\% & 2.66 & 8.25\% \\  
	\hline
	7 & 2.40 & 8.51\% & 2.38 & 8.60\%& 2.34 & 6.80\% \\ 
	\hline
	8 & 3.17 & 10.10\% & 3.31 & 11.42\%& 3.20 & 11.07\% \\
	\hline
    9 & 2.71 & 10.85\% & 2.82 & 14.33\% & 2.79 & 13.47\% \\ 
	\hline
    10 & 2.88 & 7.14\% & 2.95 & 9.53\% & 2.99 & 9.28\% \\  
	\hline
	11 & 2.06 & 9.52\% & 2.04 & 10.17\%& 2.00 & 13.85\% \\ 
	\hline
	12 & 2.50 & 13.19\% & 2.54 & 14.43\%& 2.51 & 13.45\% \\
	\hline
    13 & 2.22 & 5.21\% & 2.25 & 6.50\% & 2.21 & 5.08\% \\ 
	\hline
    14 & 2.88 & 7.50\% & 2.94 & 9.53\% & 2.93 & 9.28\% \\  
	\hline
	15 & 1.97 & 9.18\% & 1.95 & 9.22\%& 1.98 & 9.17\% \\ 
	\hline
	16 & 3.31 & 8.35\% & 3.35 & 11.07\%& 3.39 & 11.74\% \\ 
	\hline
	17 & 2.74 & 5.07\% & 2.79 & 3.40\%& 2.74 & 5.16\% \\
	\hline
    18 & 3.43 & 5.87\% & 3.50 & 8.83\% & 3.47 & 7.78\% \\ 
	\hline
    19 & 2.35 & 9.74\% & 2.43 & 13.06\% & 2.42 & 10.93\% \\  
	\hline
\end{tabular}
\end{center}
\end{table}

Table \ref{tbl:CLPU_ratio} shows the values of $R_{CLPU}$ and the LS-SVM estimation MAPE. The performance of LS-SVM has been compared with two other regression models: 1) the autoregressive model with exogenous input variables (ARX), and 2) the polynomial NARX (P-NARX) model \cite{NARX_1}. As is observed in Table \ref{tbl:CLPU_ratio}, LS-SVM shows better $R_{CLPU}$ estimation accuracy compared with the other two models. From Table \ref{tbl:outage_cases} and Table \ref{tbl:CLPU_ratio}, the impact of outage duration and ambient temperature on the ratio can be observed. 

\begin{figure}
\centering
\includegraphics[width=0.85\linewidth]{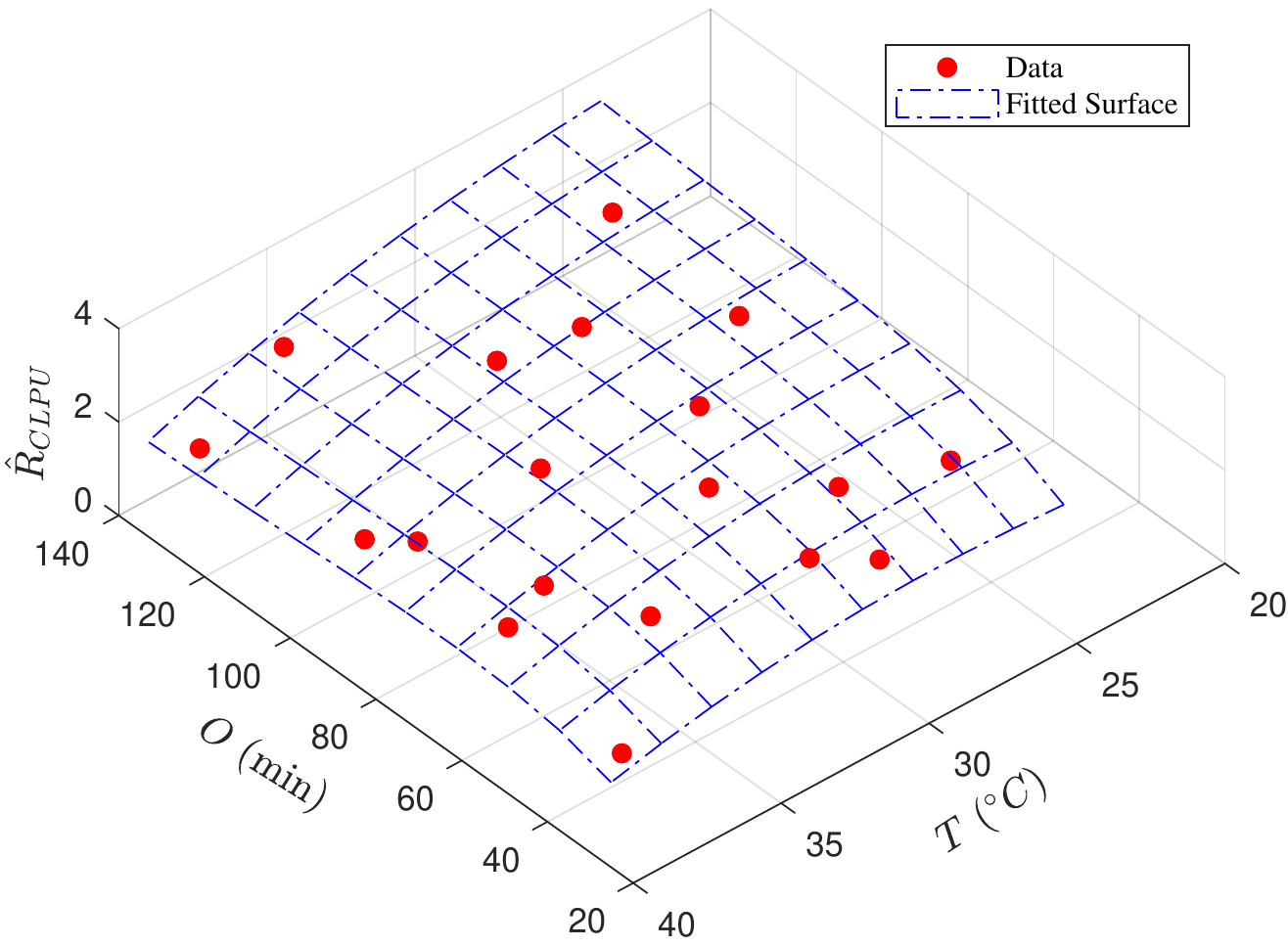}
\caption{Regression analysis of estimated CLPU ratios.}
\label{fig:CLPU_regression}
\end{figure}

Fig. \ref{fig:CLPU_regression} shows the regression analysis result of the estimated CLPU ratio in terms of outage duration ($O$) and ambient temperature ($T$). As can be seen, a surface is fitted to the data with acceptable accuracy using polynomial regression based on the estimated CLPU ratios. This CLPU ratio regression model provides an alternative way for estimating the CPLU ratio and demand in future system restoration cases. Also, as more outage cases are collected, the accuracy of the CLPU ratio regression model can be improved.

\subsubsection{Model Robustness Evaluation}
\begin{figure}
\centering
\includegraphics[width=0.85\linewidth]{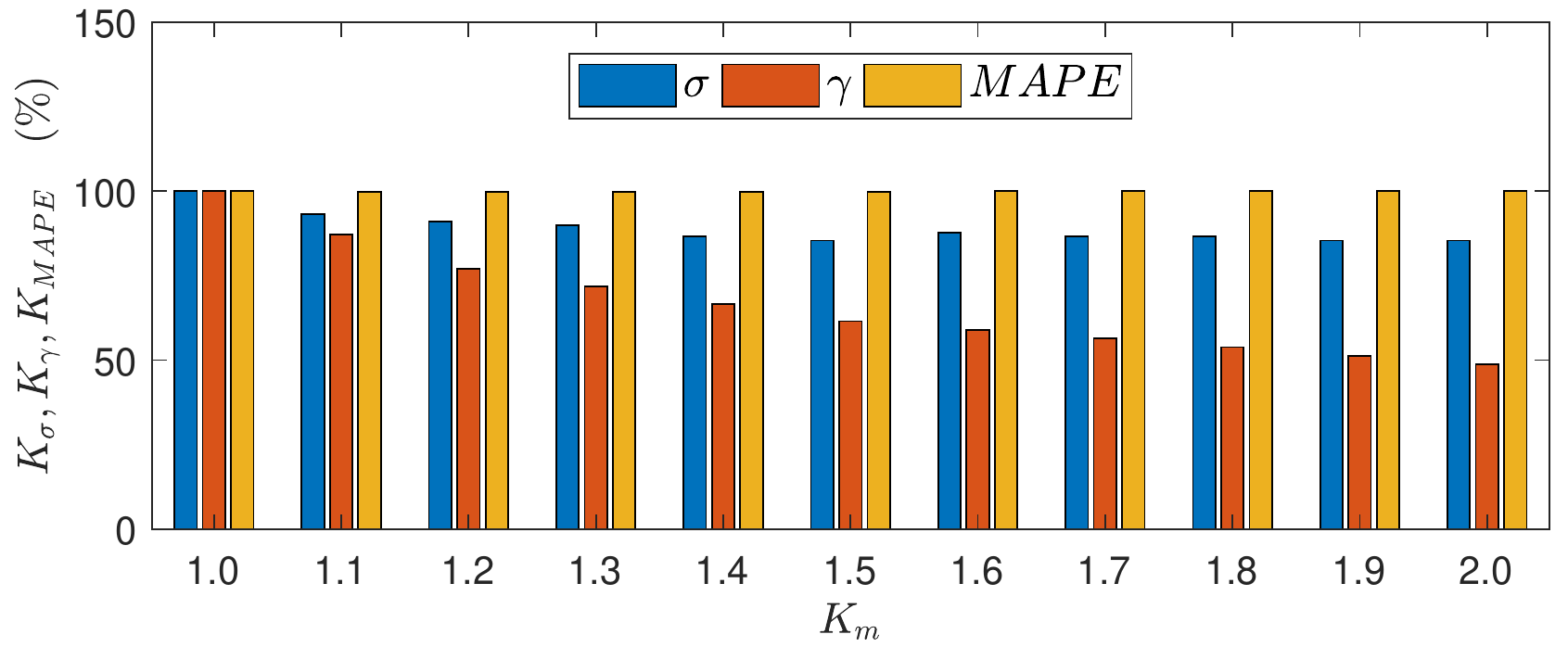}
\caption{Robustness evaluation of learning parameters.}
\label{fig:robustness}
\end{figure}

The robustness of learning parameters ($\sigma$ and $\gamma$) has been tested in accordance with \cite{LSSVM_robustness}. To do this, the following steps have been performed: first, 1\% of demand samples in the training set are randomly selected. Second, the selected samples are contaminated by multiplying them with different contamination coefficients to generate outliers. The contamination coefficient is varied from 1 to 2 with step of 0.1. The contaminated demand outliers can be written as follows:
\begin{equation}    \label{eq:outlier}
P_{ou}=P_{or}K_m
\end{equation}
where, $P_{or}$ is the original uncontaminated demand sample and $K_m$ denotes the contamination coefficient. Third, for each contamination coefficient, the LS-SVM model is retrained to obtain the new learning parameters. Also, the MAPE under these new parameters is obtained over the test set. Finally, the ratios of the retrained learning parameters (with outliers) to the original learning parameters (without outliers) are calculated to quantify the changes in the model due to noise injection. These ratios are also obtained by dividing the estimation MAPEs of the model with and without outliers. The ratios are written as follows:

\begin{equation}    \label{eq:K_sigma}
K_{\sigma}=\frac{\sigma_{ou}}{\sigma_{or}}\times 100 \%
\end{equation}

\begin{equation}    \label{eq:K_gamma}
K_{\gamma}=\frac{\gamma_{ou}}{\gamma_{or}}\times 100 \%
\end{equation}

\begin{equation}    \label{eq:K_MAPE}
K_{MAPE}=\frac{MAPE_{ou}}{MAPE_{or}}\times 100 \%
\end{equation}
where, $\sigma_{ou}$, $\gamma_{ou}$ denote the retrained learning parameters after contamination, $MAPE_{ou}$ denotes the estimation MAPE corresponding to $\sigma_{ou}$ and $\gamma_{ou}$; similarly, $\sigma_{or}$, $\gamma_{or}$ denote the trained learning parameters obtained from the original uncontaminated training set, and $MAPE_{or}$ corresponds to $\sigma_{or}$ and $\gamma_{or}$.
 Fig. \ref{fig:robustness} shows the changes in learning parameters and test MAPE against the contamination coefficient. It can be seen that the estimation MAPE has been kept unchanged, since the LS-SVM has been able to automatically adjust itself to higher levels of noise to maintain satisfactory performance. In addition, parameter $\sigma$ has been nearly kept constant as the contamination coefficient increases. However, parameter $\gamma$ decreases as the contamination coefficient increases, which can be justified using equation \eqref{eq:LSSVM_eq3}. As can be seen, in equation \eqref{eq:LSSVM_eq3}, $\gamma$ is basically the regularization trade-off factor which determines the weight of the training set's inherent noise level during the training process. Hence, when we artificially increase this noise level in the training set through contamination, the LS-SVM training algorithm automatically decreases the weight assigned to the inherent noise parameter to keep the risk function at its minimum value.

\begin{figure}
\centering
\subfloat[Actual CLPU ratios.\label{sfig:actual}]{
\includegraphics[width=0.8\linewidth]{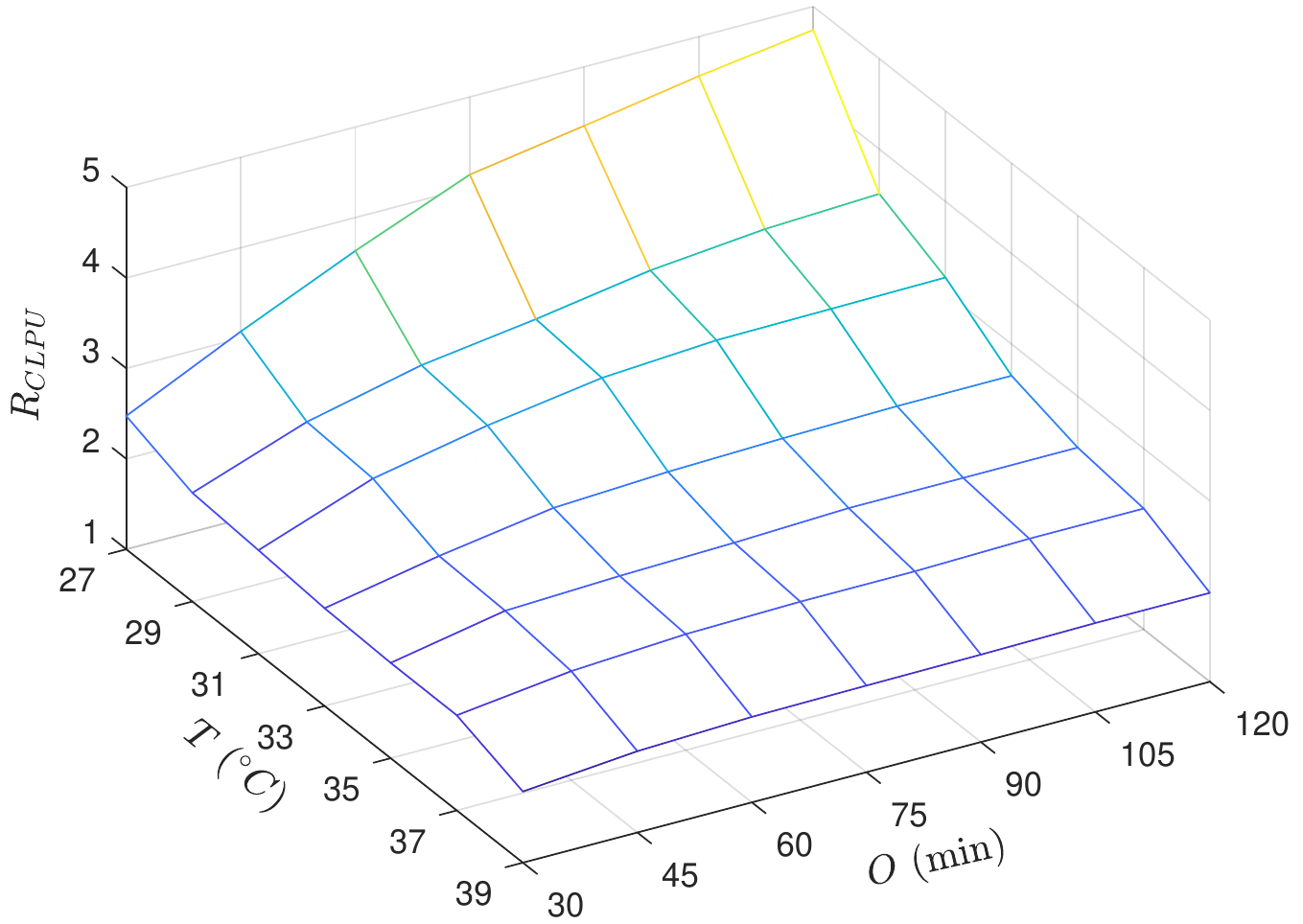}
}
\hfill
\subfloat[Estimated CLPU ratios.\label{sfig:estimated}]{
\includegraphics[width=0.8\linewidth]{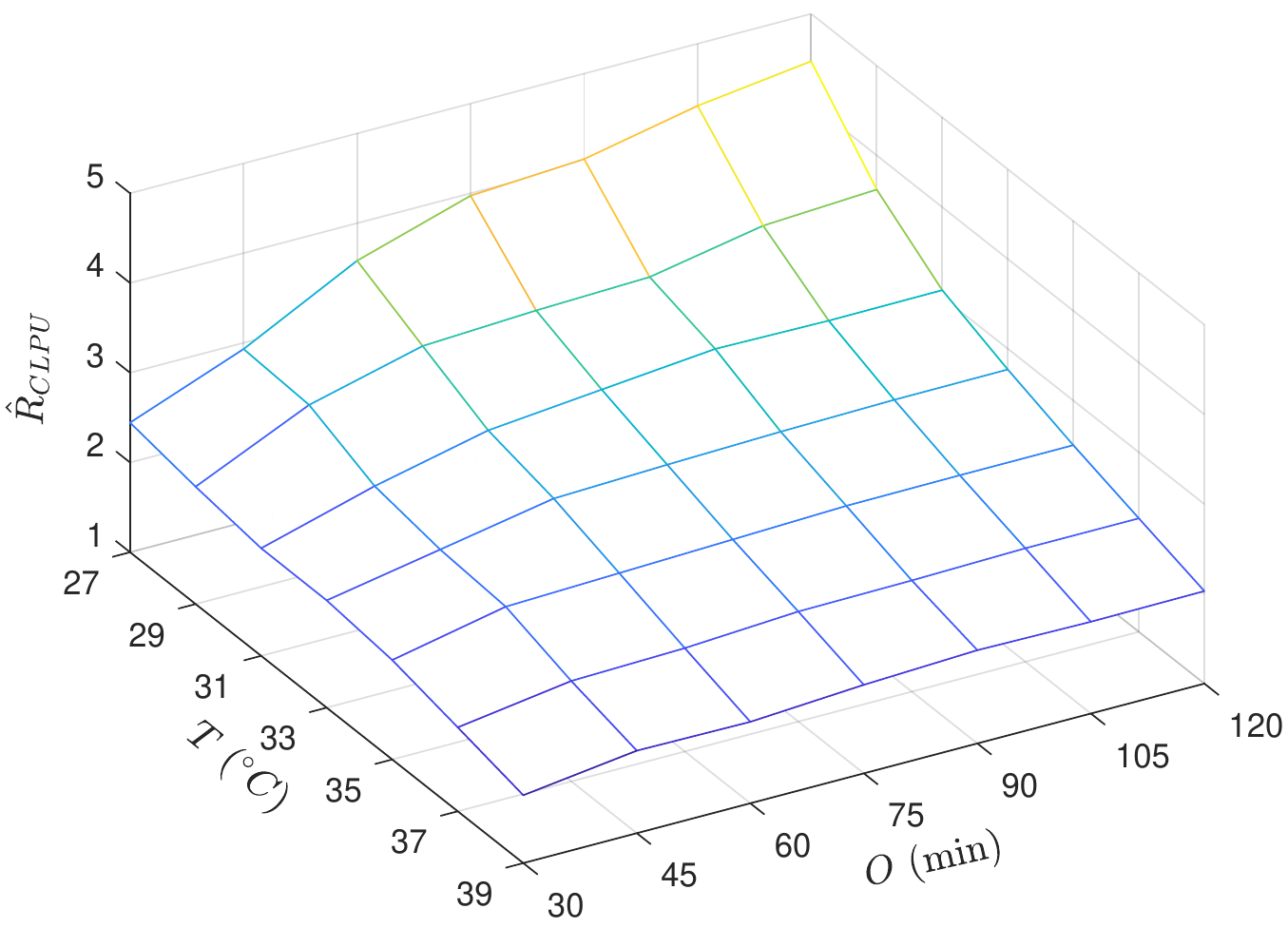}
}
\caption{Actual and estimated CLPU ratios.}
\label{fig:actual_estimated_CLPU_ratio}
\end{figure}
\subsubsection{CLPU Ratio Validation}
\begin{figure}
\centering
\includegraphics[width=0.80\linewidth]{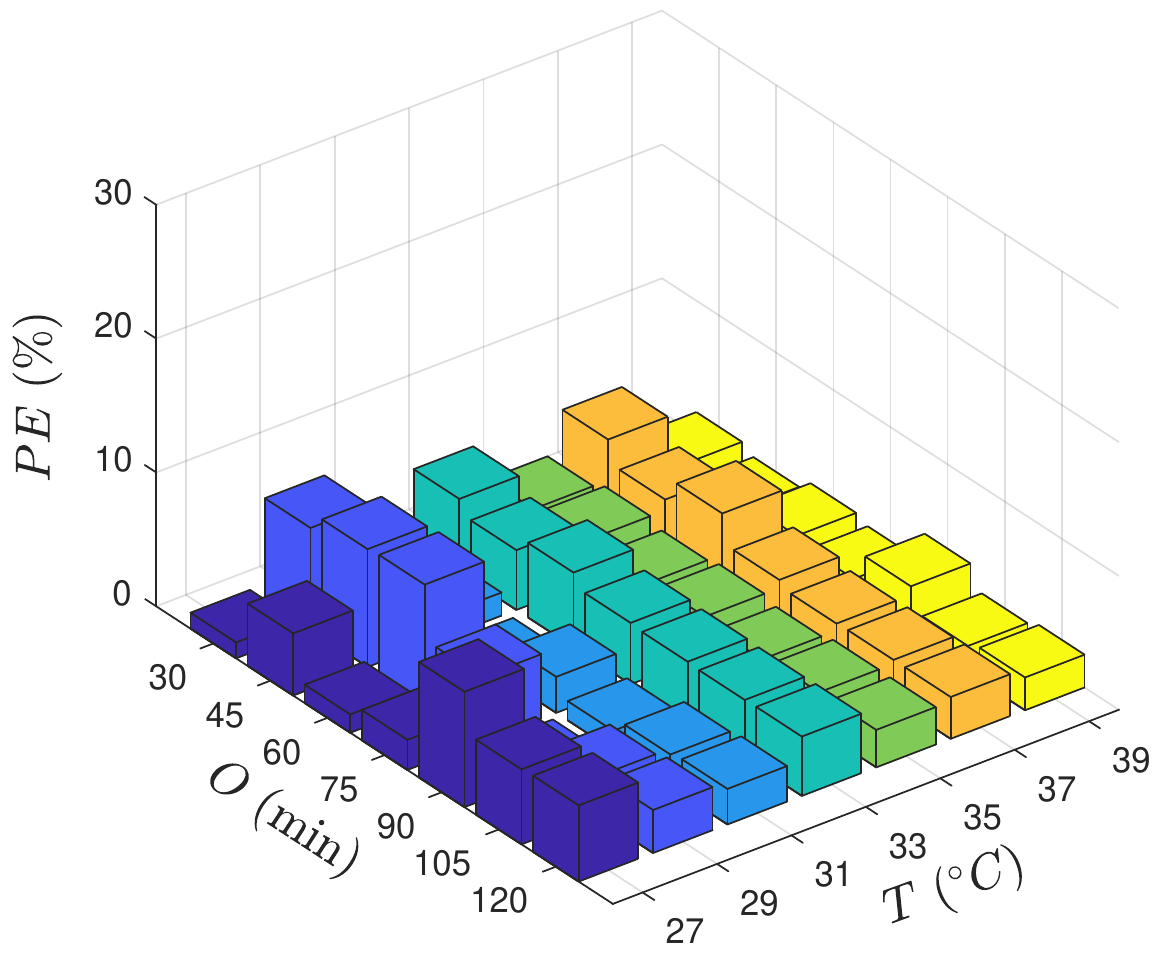}
\caption{CLPU ratio estimation percentage errors.}
\label{fig:PE_RCLPU}
\end{figure}

Since the real diversified demand at the time of restoration is unknown, due to the undiversified nature of load, therefore, we cannot validate the estimated ratios using real outage cases alone. Considering this, we have conducted additional Monte Carlo simulations to validate the CLPU ratios. Specifically, 49 new outage cases with different outage durations and different ambient temperatures are created, and then, our proposed approach is applied to the data generated from these outage cases \cite{TCL_model}. The basic steps of validation are as follows: first, the demand consumed by TCLs are generated using Monte Carlo simulations for a heterogeneous population of customers, obtained from appliance-level state-space modeling, both in normal operation and outage conditions; second, additional appliances' consumed demands are added to the TCLs' consumed demands to obtain customers' net demands; third, feeder-level demand is obtained by aggregating customer-level loads; then, the proposed CLPU assessment framework is applied to the data obtained from 49 outage cases to obtain corresponding CLPU ratios; finally, ratio validation is conducted by comparing the outcomes of the proposed data-driven model and the simulation results. Fig. \ref{sfig:actual} and Fig. \ref{sfig:estimated} show the actual and estimated CLPU ratios, respectively. Note that the actual CLPU ratios are obtained from Monte Carlo simulations, and the estimated CLPU ratios are obtained by applying our proposed framework to the demand data generated from these Monte Carlo simulations. In Fig. \ref{sfig:actual} and Fig. \ref{sfig:estimated}, $T$  denotes the ambient temperature and $O$ denotes outage duration. As can be seen, the estimated CLPU ratios can accurately match the actual CLPU ratios. The validation of CLPU ratio can also be demonstrated in Fig. \ref{fig:PE_RCLPU}, in which the CLPU ratio estimation percentage errors ($PE$) are smaller than 9$\%$ for all cases, and 90\% of the percentage error values are less than 6\%, which validates the performance of the framework. This can also demonstrate the advantage of our proposed data-driven approach over the model-driven Monte Carlo simulator, showing that the CLPU ratio can be accurately estimated only based on the available demand data and without the knowledge of thermal parameters of individual customer houses.

\begin{figure}
\centering
\includegraphics[width=0.73\linewidth]{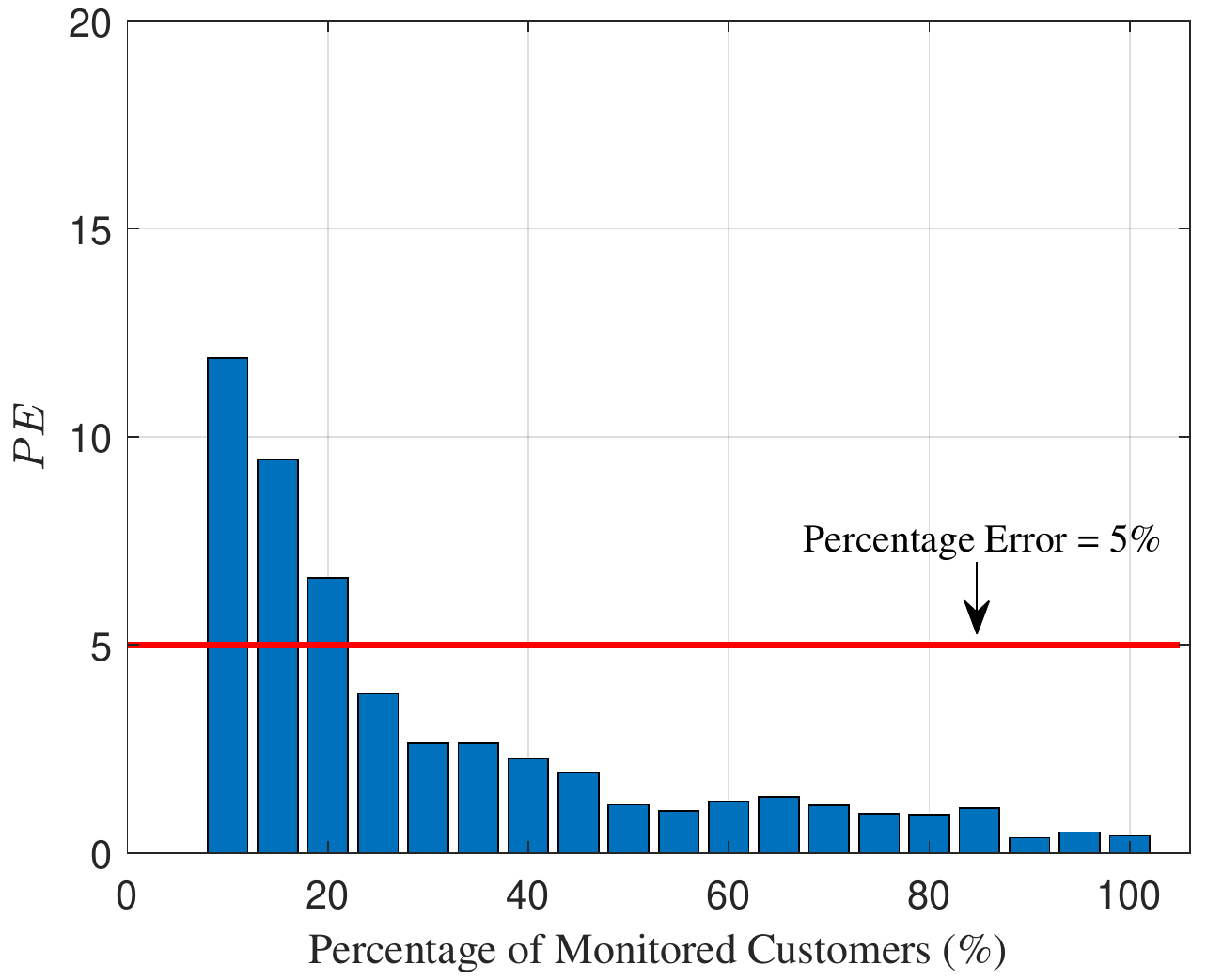}
\caption{Relationship between the CLPU ratio estimation percentage error with the percentage of monitored customers.}
\label{fig:PE_RCLPU_PE_cust}
\end{figure}

In practice, it is probable that a proportion of customers are unmonitored. Hence, it is of interest to analyze the performance of the proposed framework in scenarios where different proportions of customers do not have smart meters. This has also been demonstrated using Monte Carol simulations, where the CLPU ratio estimation percentage errors are shown as a function of percentage of monitored customers in Fig. \ref{fig:PE_RCLPU_PE_cust}. It can be seen that as the number of monitored customers increases the accuracy of the proposed framework improves. Also, the framework still has acceptable accuracy even when a high percentage of customers are unmonitored.

\begin{figure}
\centering
\includegraphics[width=0.75\linewidth]{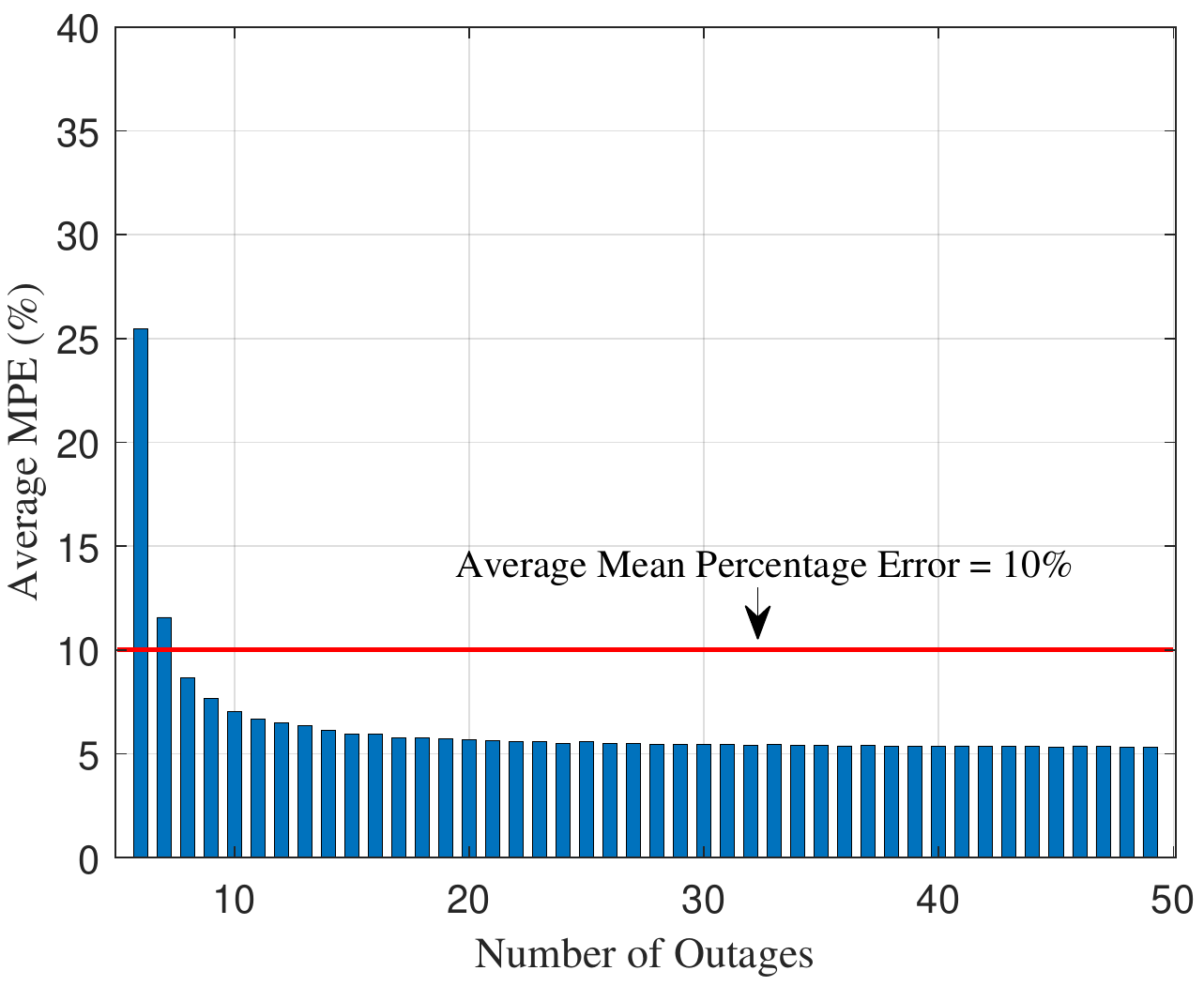}
\caption{Relationship between average MPE with the number of outages.}
\label{fig:robustness_num_outages}
\end{figure}

It is also of great importance to conduct robustness analysis with respect to the number of available outage cases due to the outage data scarcity. To do this, a training process has been performed using random drop-out for cross validation. The performance of the framework has been evaluated in terms of the CLPU ratio prediction mean percentage error (MPE), and is plotted against the number of historical outages, as shown in Fig. \ref{fig:robustness_num_outages}. As can be seen, to reach an average MPE of smaller than 10\%, a minimum number of eight outages is required in this case. Hence, as more outage data become available, the accuracy of the regression model is improved. This robustness analysis has also been conducted on our utility data, and a similar decreasing trend of average MPE against the number of outages is observed.

\subsection{Customer CLPU Demand Increase Estimation}
\begin{figure}
\centering
\subfloat[Empirical histogram\label{sfig:j1}]{
\includegraphics[width=0.8\linewidth]{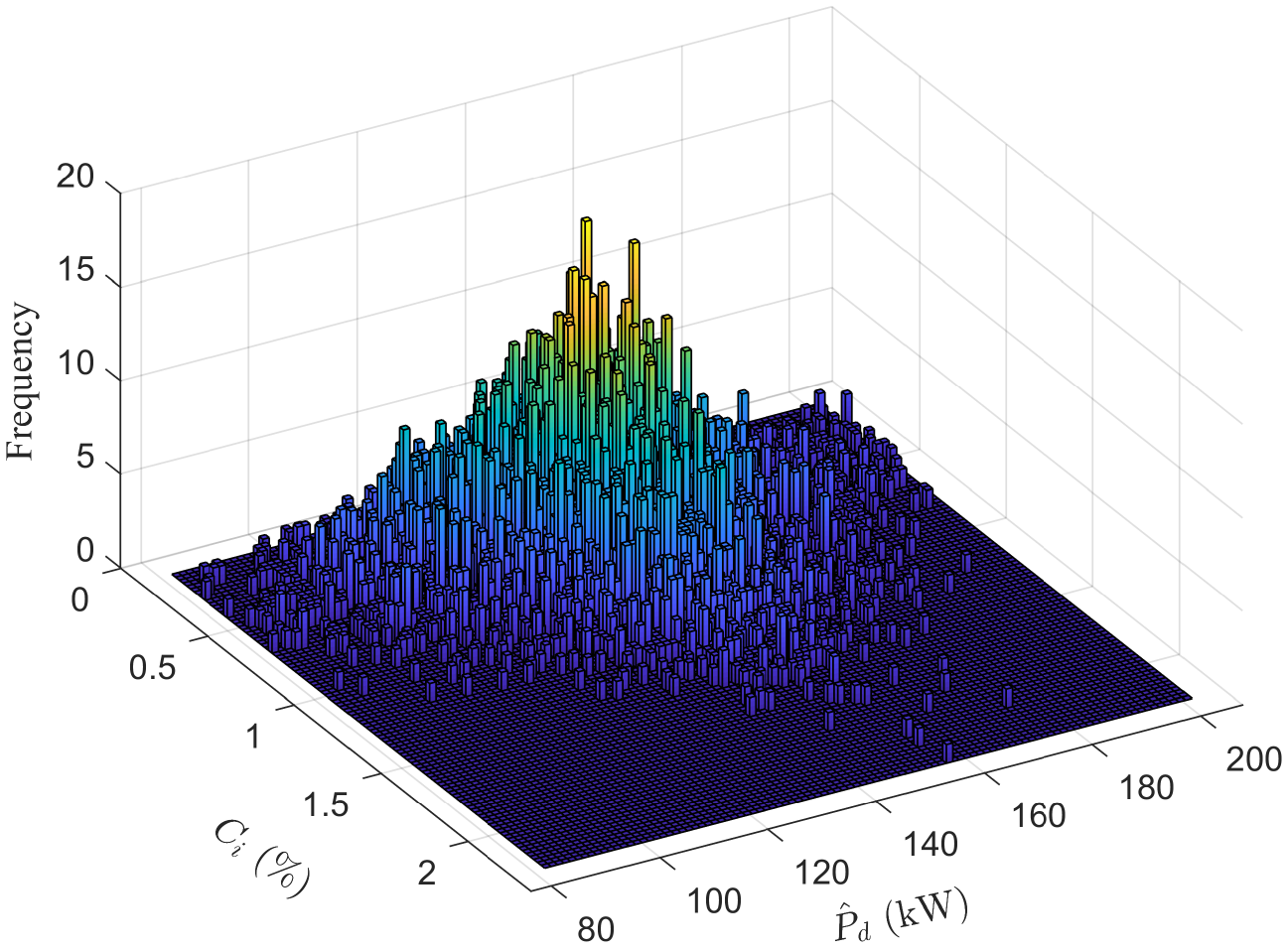}
}
\hfill
\subfloat[GMM-based estimation\label{sfig:j2}]{
\includegraphics[width=0.8\linewidth]{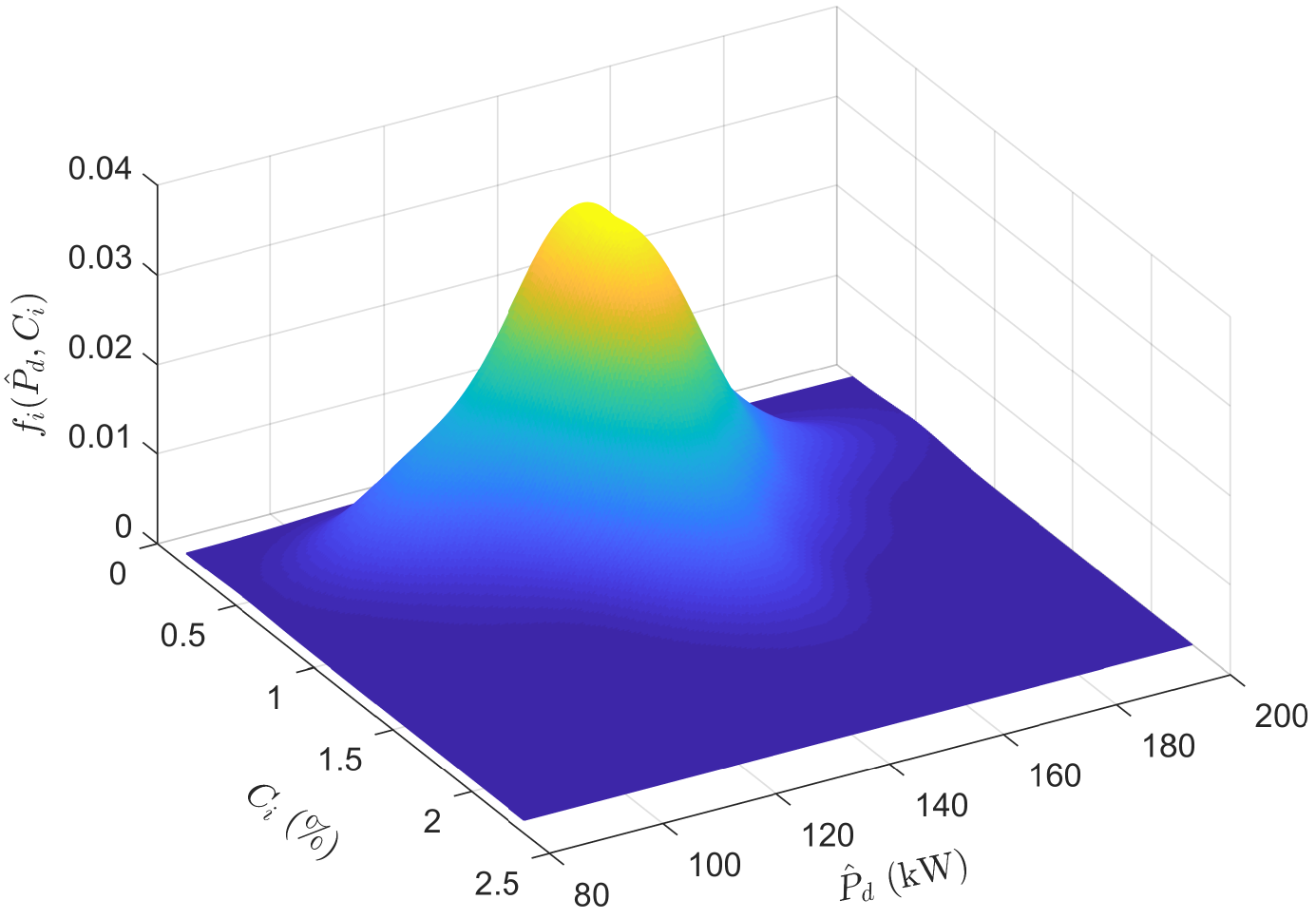}
}
\caption{Joint PDF estimation of diversified feeder demand and contribution factor for one customer.}
\label{fig:joint}
\end{figure}

\begin{figure}[h]
\centering
\subfloat[Distribution of $\hat{p}_{d,i}$ \label{sfig:distri_p_hat}]{
\includegraphics[width=0.8\linewidth]{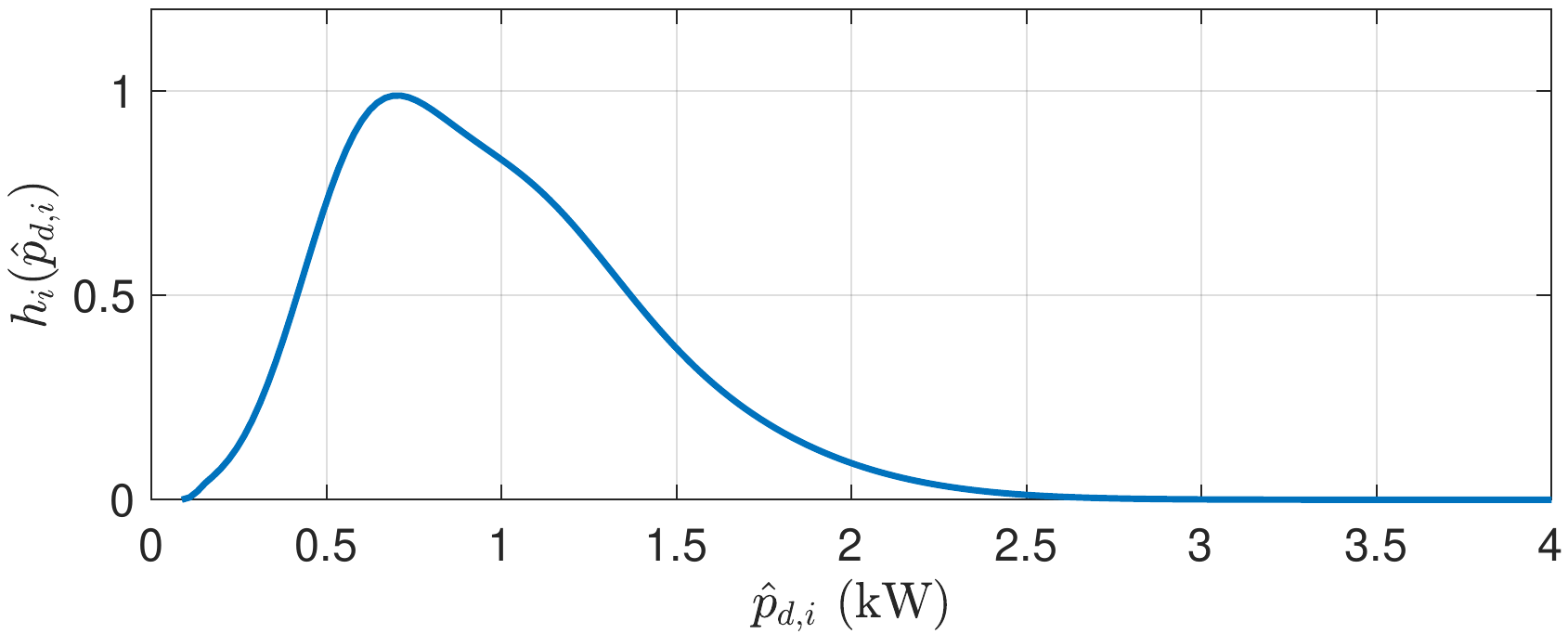}
}
\hfill
\subfloat[Distribution of $\hat{I}_i$ \label{sfig:distri_p_incr_hat}]{
\includegraphics[width=0.8\linewidth]{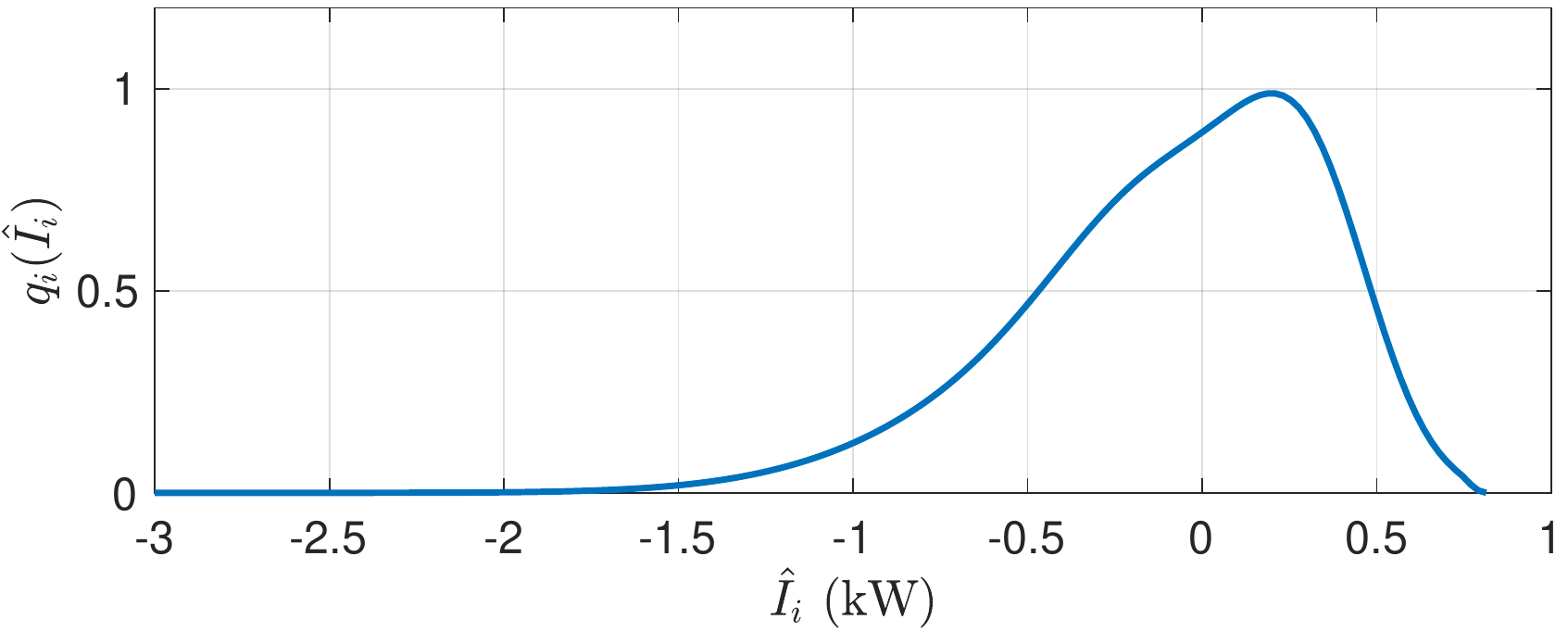}
}
\caption{Distribution of estimated demand and CLPU demand increase of one customer.}
\label{fig:cust_D_Dincre}
\end{figure}

\begin{figure}
\centering
\includegraphics[width=0.8\linewidth]{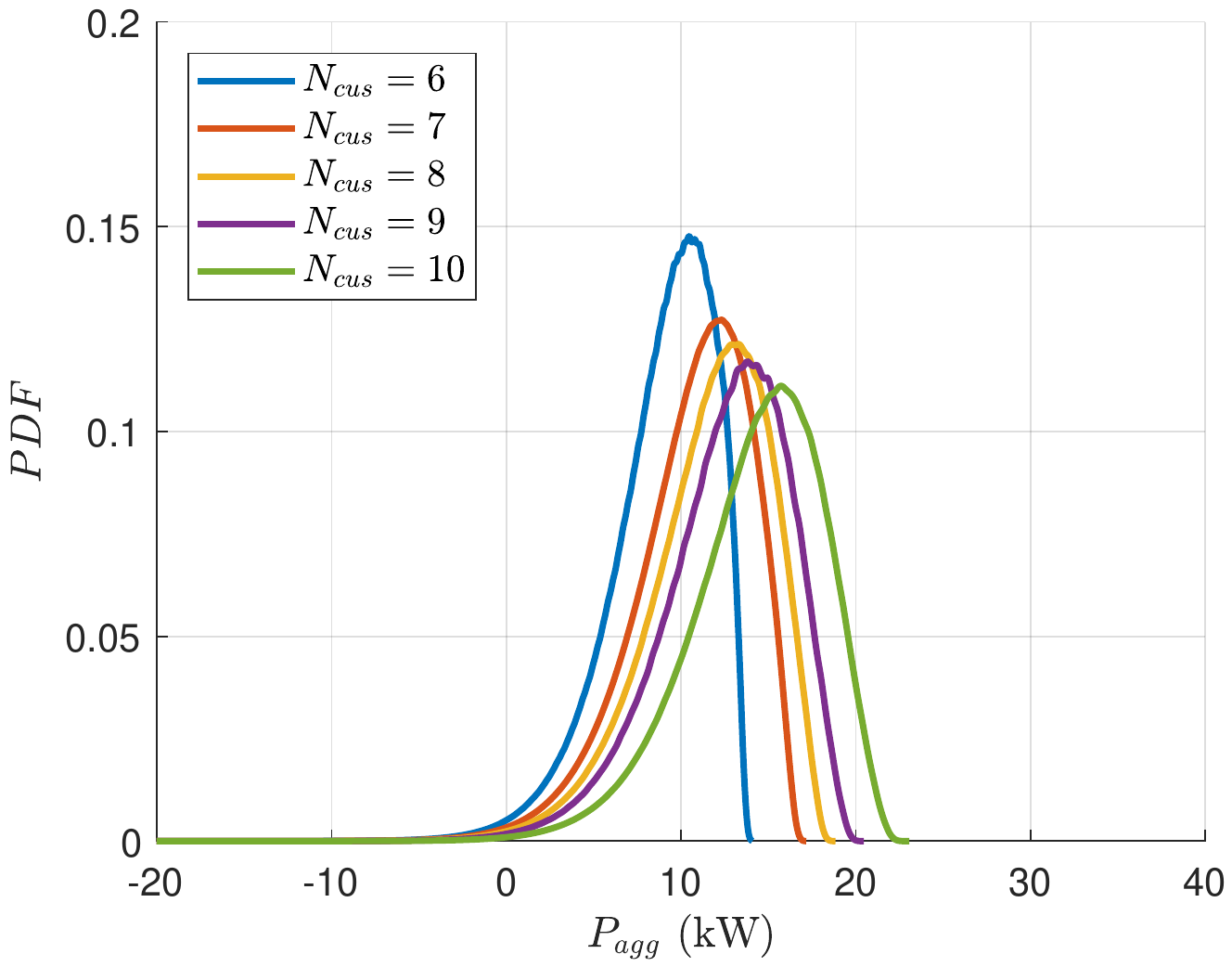}
\caption{Distributions of aggregate demand increase.}
\label{fig:P_aggre_PDF}
\end{figure}

Fig. \ref{sfig:j1} and Fig. \ref{sfig:j2} show the empirical histogram and the GMM-based estimation of $f_i(\hat{P}_d,C_i)$ for one customer, respectively. As can be seen by comparing these figures, GMM is able to accurately model the behavior of the customer using smooth parametric Gaussian density functions. 
Fig. \ref{fig:cust_D_Dincre} shows the probability distribution of estimated demand and CLPU demand increase of one customer at time $t_r$. Note that the probable CLPU demand increase of the customer can be negative. This partly reflects the stochasticity of customer demand. Regarding system restoration issue, optimal approaches have been proposed for restoring different groups of customers after extreme events in the literature \cite{restore_3,restore_4,restore_5}. Our proposed method can provide the marginal PDF of demand increase for a group of customers by convolving the marginal PDFs of demand increase of individual customers \cite{convolution}, which is useful for the utilities to perform restoration risk evaluation. For instance, Fig. \ref{fig:P_aggre_PDF} shows the PDFs of aggregate demand increase ($P_{agg}$) for $N_{cus}$ customers connected to the same transformer. Hence, the impact of CLPU demand increase on the transformer can be accurately quantified. As can be seen, as the number of customers increases the expected aggregate demand increase also shifts towards larger values.

To evaluate the loss of load diversity for a population of customers, the following index is defined for each customer:
\begin{equation}  \label{eq:cust_7}
\mathbb{P}_{I_0,i}=\Pr(\hat{I}_i\ge I_0)
\end{equation}
where, $\mathbb{P}_{I_0,i}$ denotes the probability of estimated demand increase being larger than a threshold, $I_0$, for the $i^{th}$ customer, with $\Pr(a)$ defining probability of event $a$. Using this index, the factor $R_{lb}(I_0)$ indicates the percentage of customers with $\mathbb{P}_{I_0,i} > 0$, as shown in equation \eqref{eq:cust_8}:
\begin{equation}  \label{eq:cust_8}
R_{lb}(I_0)=\frac{\sum_{i=1}^{M}H(\mathbb{P}_{I_0,i})}{M}\times 100\%
\end{equation}
where, $H(x)$ is the Heaviside step function defined as follows:
\begin{equation}  \label{eq:cust_9}
H(x)=\begin{cases}
1 & x\ge 0 \\
0 & x<0
\end{cases}
\end{equation}

\begin{figure}
\centering
\subfloat[The relationship between $R_{lb}$ and $I_0$\label{sfig:Rlb}]{
\includegraphics[width=0.8\linewidth]{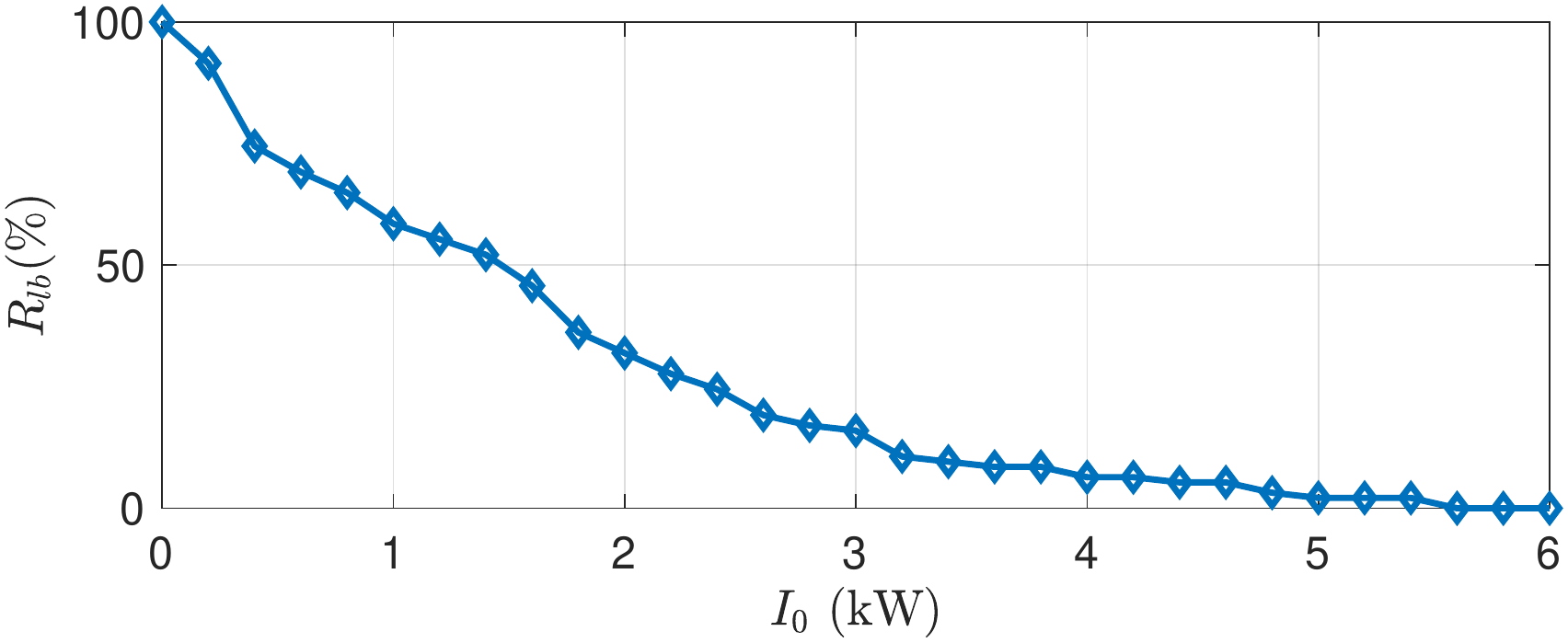}
}
\hfill
\subfloat[Distribution of $\mathbb{P}_{I_0,i}$ \label{sfig:boxplot}]{
\includegraphics[width=0.8\linewidth]{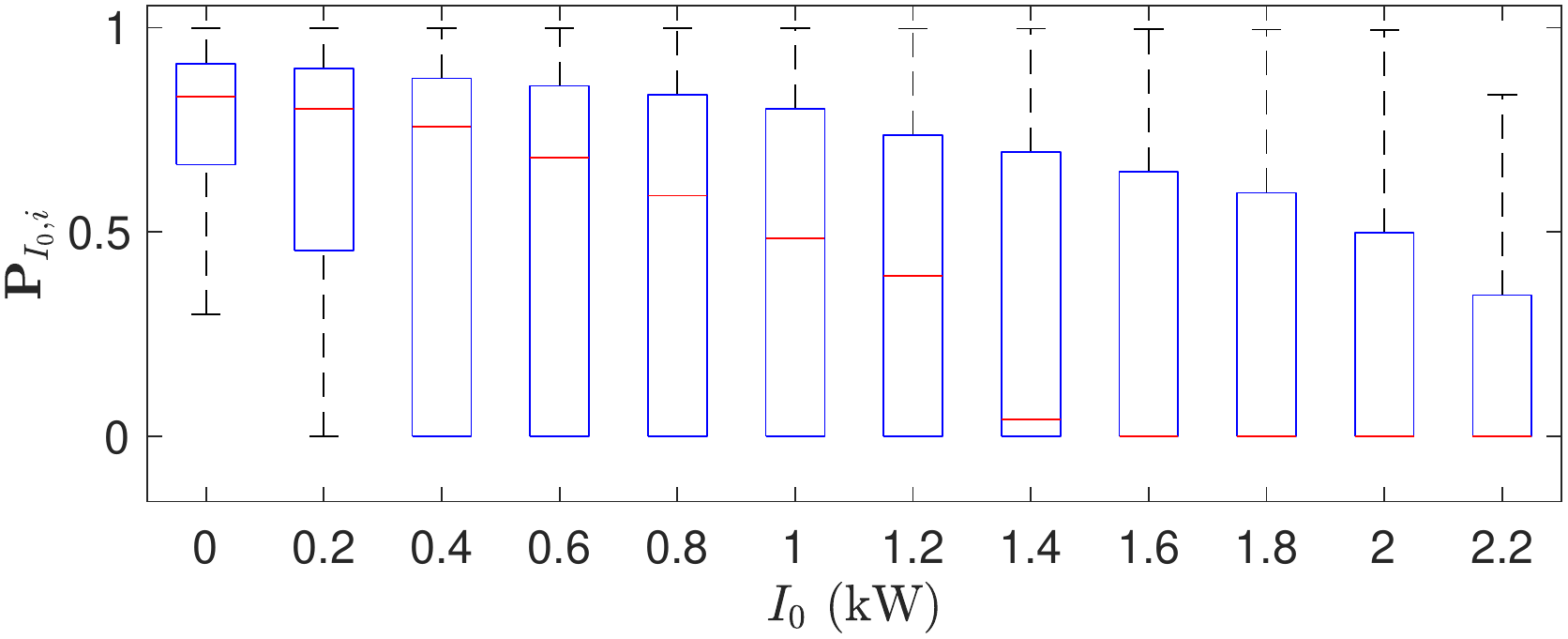}
}
\caption{Evaluation of loss of load diversity.}
\label{fig:loss_diversity}
\end{figure}

Fig. \ref{fig:loss_diversity}a shows the relationship between $R_{lb}$ and $I_{0}$. It can be seen that: 1) for $I_0 = 0$ we have $R_{lb} = 100\%$, which implies that all customers have non-negative CLPU demand increase with non-zero probability, and 2) $R_{lb}$ decreases as $I_{0}$ increases, which indicates that the number of customers with $\hat{I}_i > I_0$ decreases as $I_0$ increases. This is determined by the maximum capability of customers' contribution to feeder CLPU demand.
Fig. \ref{fig:loss_diversity}b shows $\mathbb{P}_{I_0,i}$ distribution boxplot as a function of threshold level $I_0$. This figure describes the loss of load diversity during service restoration. For example, the first box tells us that almost all of the customers have $\mathbb{P}_{I_0 = 0,i} > 0.5$ for this outage case. This means that nearly all customer' loads simultaneously start drawing more energy than normal from the feeder in the restoration phase. It can also be seen that the first quartile, the median, and the third quartile values of $\mathbb{P}_{I_0,i}$ present a descending trend as the threshold $I_{0}$ increases. This is consistent with the decreasing trend of $R_{lb}$, observed in Fig. \ref{fig:loss_diversity}a. This implies that only a few customers have abnormaly high demand increase during service restoration.

\begin{figure}[h]
\centering
\includegraphics[width=0.75\linewidth]{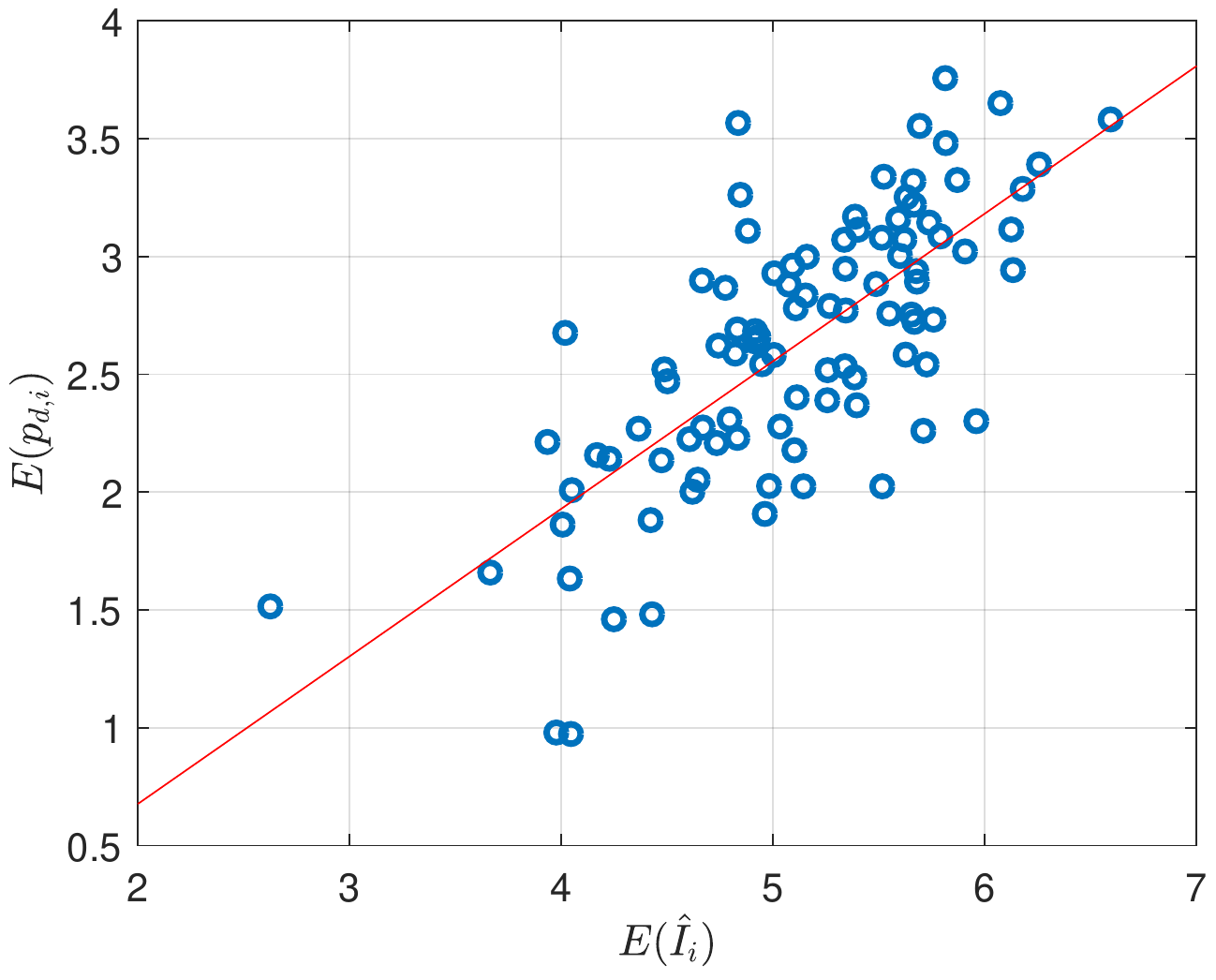}
\caption{Relationship between entropy of customer CLPU demand increase and entropy of normal demand at the time corresponding to restoration instant.}
\label{fig:entropy}
\end{figure}

It is also of interest to discover the relationship between the uncertainty of customer demand increase and the uncertainty of customer behavior during normal system operation. To evaluate the uncertainty of customer demand increase at the time of restoration, the \textit{entropy} of $\hat{I}_i$ is obtained using $q_i(\hat{I}_i)$, as follows \cite{entropy_definition}:
\begin{equation}  \label{eq:df_entropy}
E(\hat{I}_i)=-\int_{\hat{I}_i}q_i(\hat{I}_i) \log_2(q_i(\hat{I}_i)) d\hat{I}_i
\end{equation}
On the other hand, to evaluate the uncertainty of customer behavior during normal system operation, customer demand is sampled on different days at the same time corresponding to the restoration instant. Based on these data samples the entropy of customer behavior is defined similar to \eqref{eq:df_entropy} and denoted as $E(p_{d,i})$. Fig. \ref{fig:entropy} shows the relationship between $E(\hat{I}_i)$ and $E(p_{d,i})$ for all customers. It can be seen that a positive linear relationship exists between the uncertainty of customer CLPU demand increase and the uncertainty of normal customer demand at the time corresponding to the restoration instant. The correlation between these two entropy variables is around 0.72, which implies that customers with uncertain normal demand also show more uncertainty at the time of restoration.

\begin{figure}[h]
\centering
\includegraphics[width=0.75\linewidth]{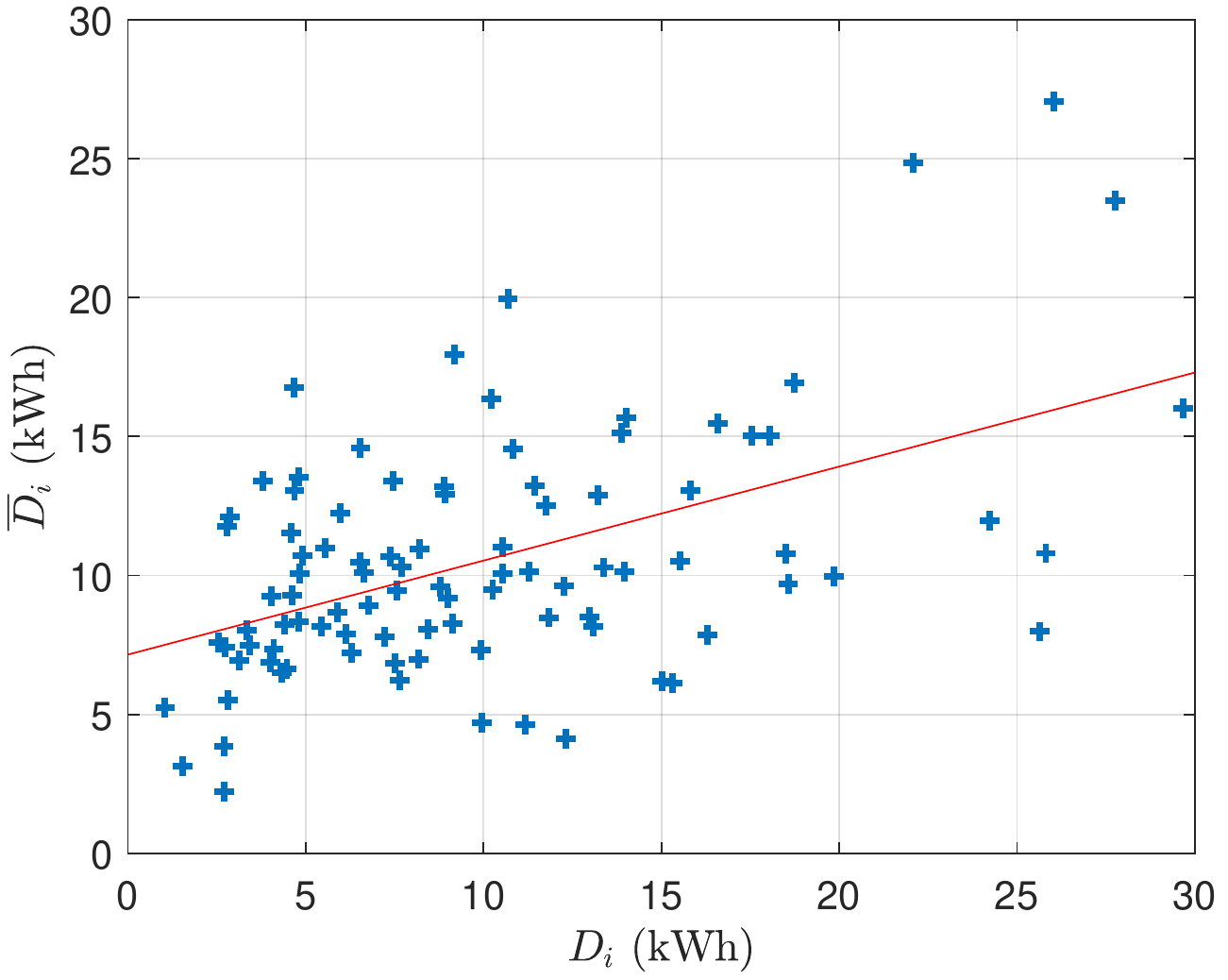}
\caption{Relationship between customer energy consumption after the time of restoration and average normal energy consumption at corresponding time period.}
\label{fig:energy_increase}
\end{figure}

To assess customer demand increase due to CLPU, the relationship between energy consumption within a 4-hour time interval after time of restoration, $D_i$, and average energy consumption during the same time period in normal operation, $\overline{D}_i$, is analyzed. Fig. \ref{fig:energy_increase} shows the relationship between $D_i$ and $\overline{D}_i$ for all customers. The slope of the fitted line is smaller than 1, which indicates that customer energy consumption after the time of restoration is greater than average energy consumption in normal operation during the corresponding time period. Also, a positive correlation between $D_i$ and $\overline{D}_i$ is observed, which implies that higher energy consumption during normal system operation corresponds to higher restoration energy consumption.

\section{Conclusion}\label{sec:conclusion}
This paper has presented a data-driven framework for using smart meter data to determine feeder-level CLPU demand ratio and to assess customer-level demand increase due to CLPU, based on historical outage cases. Machine learning and probabilistic methodologies are used for CLPU demand assessment. Outage cases are employed for model training and verification. The results of case studies show that the proposed framework can accurately determine feeder-level CLPU demand ratio and assess customer-level demand increase due to loss of load diversity during service restoration. It is shown that only a few customers have extreme CLPU demand increase, and customers with higher energy consumption during normal operation typically have higher demand during the restoration phase. The performance of the proposed data-driven framework is validated using extensive Monte Carlo simulations. It has been demonstrated that our method is able to accurately assess CLPU demand at both feeder- and customer-levels without having any explicit knowledge of individual houses' thermal information.


\ifCLASSOPTIONcaptionsoff
  \newpage
\fi



\bibliographystyle{IEEEtran}
\bibliography{IEEEabrv,./bibliographies/CLPU}   

\begin{IEEEbiography}[{\includegraphics[width=1in,height=1.25in,clip,keepaspectratio]{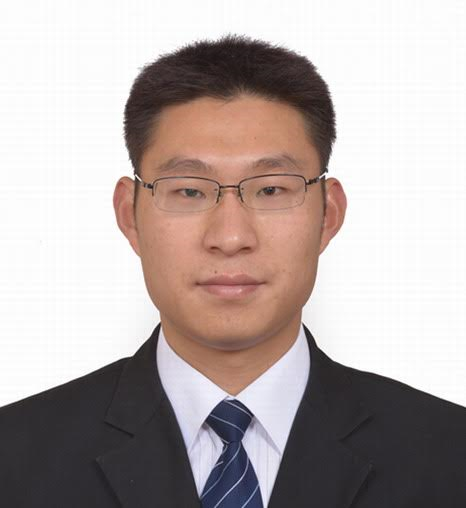}}]{Fankun Bu}(S'18) received the B.S. and M.S. degrees from North China Electric Power University, Baoding, China, in 2008 and 2013, respectively. From 2008 to 2010, he worked as a commissioning engineer for NARI Technology Co., Ltd., Nanjing, China. From 2013 to 2017, he worked as an electrical engineer for State Grid Corporation of China at Jiangsu, Nanjing, China. He is currently pursuing his Ph.D. in the Department of Electrical and Computer Engineering, Iowa State University, Ames, IA. His research interests include load modeling, load forecasting, data analytics in distribution system, and power system relaying.
\end{IEEEbiography}
\begin{IEEEbiography}[{\includegraphics[width=1in,height=1.25in,clip,keepaspectratio]{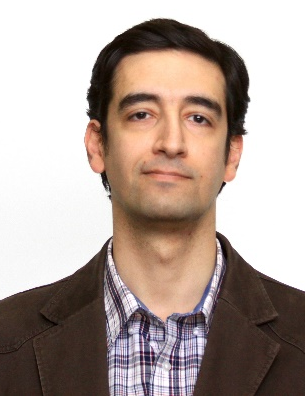}}]{Kaveh Dehghanpour}(S'14--M'17) received his B.Sc. and M.S. from University of Tehran in electrical and computer engineering, in 2011 and 2013, respectively. He received his Ph.D. in electrical engineering from Montana State University in 2017. He is currently a postdoctoral research associate at Iowa State University. His research interests include application of machine learning and data-driven techniques in power system monitoring and control.
\end{IEEEbiography}
\begin{IEEEbiography}[{\includegraphics[width=1in,height=1.25in,clip,keepaspectratio]{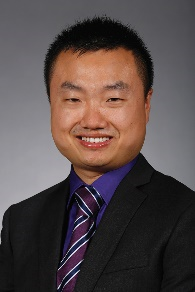}}]{Zhaoyu Wang}(S'13--M'15) is the Harpole-Pentair Assistant Professor with Iowa State University. He received the B.S. and M.S. degrees in electrical engineering from Shanghai Jiaotong University in 2009 and 2012, respectively, and the M.S. and Ph.D. degrees in electrical and computer engineering from Georgia Institute of Technology in 2012 and 2015, respectively. He was a Research Aid at Argonne National Laboratory in 2013 and an Electrical Engineer Intern at Corning Inc. in 2014. His research interests include power distribution systems, microgrids, renewable integration, power system resilience, and data-driven system modeling. He is the Principal Investigator for a multitude of projects focused on these topics and funded by the National Science Foundation, the Department of Energy, National Laboratories, PSERC, and Iowa Energy Center. Dr. Wang is the Secretary of IEEE Power and Energy Society Award Subcommittee. He is an editor of IEEE Transactions on Power Systems, IEEE Transactions on Smart Grid and IEEE PES Letters, and an associate editor of IET Smart Grid.
\end{IEEEbiography}
\begin{IEEEbiography}[{\includegraphics[width=1in,height=1.25in,clip,keepaspectratio]{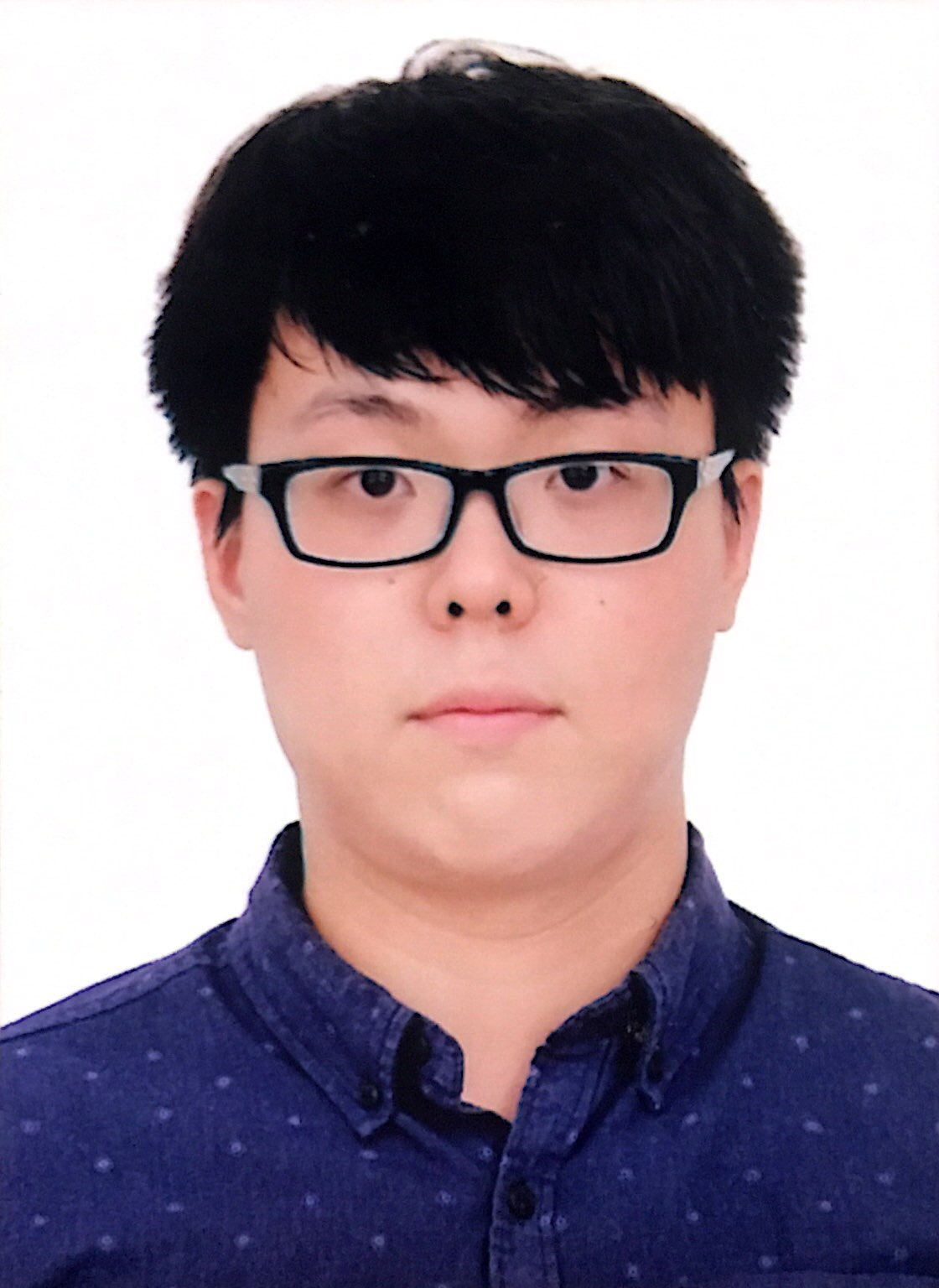}}]{Yuxuan Yuan}(S'18)  received the B.S. degree in Electrical \& Computer Engineering from Iowa State University, Ames, IA, in 2017. He is currently pursuing the Ph.D. degree at Iowa State University. His research interests include distribution system state estimation, synthetic networks, data analytics, and machine learning.
\end{IEEEbiography}

\end{document}